\documentclass[twocolumn,showpacs,english,prb,superscriptaddress,preprintnumbers,amsmath,amssymb,floatfix,eqsecnum]{revtex4-1}
\usepackage{amssymb}
\usepackage{amsmath}
\usepackage{amsfonts}
\usepackage{dsfont}
\usepackage{bm} 
\usepackage{graphicx}
\usepackage{slashed}

\newcommand{\bk}{{\bf k}}
\newcommand{\bb}{{\bf b}}

\newcommand{\bx}{{\bf x}}
\newcommand{\by}{{\bf y}}
\newcommand{\bp}{{\bf p}}
\newcommand{\sn}{\sigma^{\mathrm{n}}}
\newcommand{\boldgreek}[1]{\mbox{\boldmath$\mathbf{#1}$\unboldmath}}
\newcommand{\bsigma}{\boldgreek{\sigma}}
\newcommand{\sla}[1]{\slashed{#1}}

\begin{document}

\title{Theory of the strongly disordered Weyl semimetal}

\author{Alexander Altland and Dmitry Bagrets}
\affiliation{Institut f\"ur Theoretische Physik, 
Universit\"at zu K\"oln, Z\"ulpicher Str.~77, D-50937 K\"oln, Germany }

\date{\today}

\begin{abstract}

In disordered Weyl semimetals, mechanisms of topological origin lead to
novel mechanisms of transport, which manifest themselves in unconventional types of electromagnetic response. Prominent examples of transport phenomena particular to the Weyl context include the anomalous Hall effect, the chiral magnetic effect, and the formation of totally field dominated regimes of transport in which the longitudinal conductance is proportional to an external magnetic field. In this paper, we discuss the manifestations of these phenomena at large length scales including the cases of strong disorder and/or magnetic field which are beyond the scope of diagrammatic perturbation theory. Our perhaps most striking finding is the identification of a novel regime of drift/diffusion transport where diffusion at short scales gives way to effectively ballistic dynamics at large scales, before a re-entrance to diffusion takes place at yet larger scales. We will show that this regime plays a key role in understanding the interplay of the various types of magnetoresponse of the system. Our results are obtained by describing the strongly disordered system in terms of an effective field theory of Chern-Simons type. The paper contains a self-contained derivation of this theory, and a discussion of both equilibrium and non-equilibrium (noise) transport phenomena following from it. 
\end{abstract}

\pacs{75.47.-m, 03.65.Vf, 73.43.-f}

\maketitle

\section{Introduction}

Topological metals are the gapless cousins of topological insulators. Where the latter support a gapped spectrum, and gap closure means a topological phase transition, the former have a gapless spectrum, and gap opening requires a phase transition. The gaplessness of the topological metal is protected by topological charges in the Brillouin zone, i.e. points or more generally submanifolds, to which topological invariants may be assigned.  Salient features characterizing the physics of topological metals include unconventional transport characteristics, protection against Anderson localization, or the appearance of unconventional structures in surface Brillouin zones (Fermi arcs).

\begin{figure}[t]
  \centering
\includegraphics[width=8.5cm]{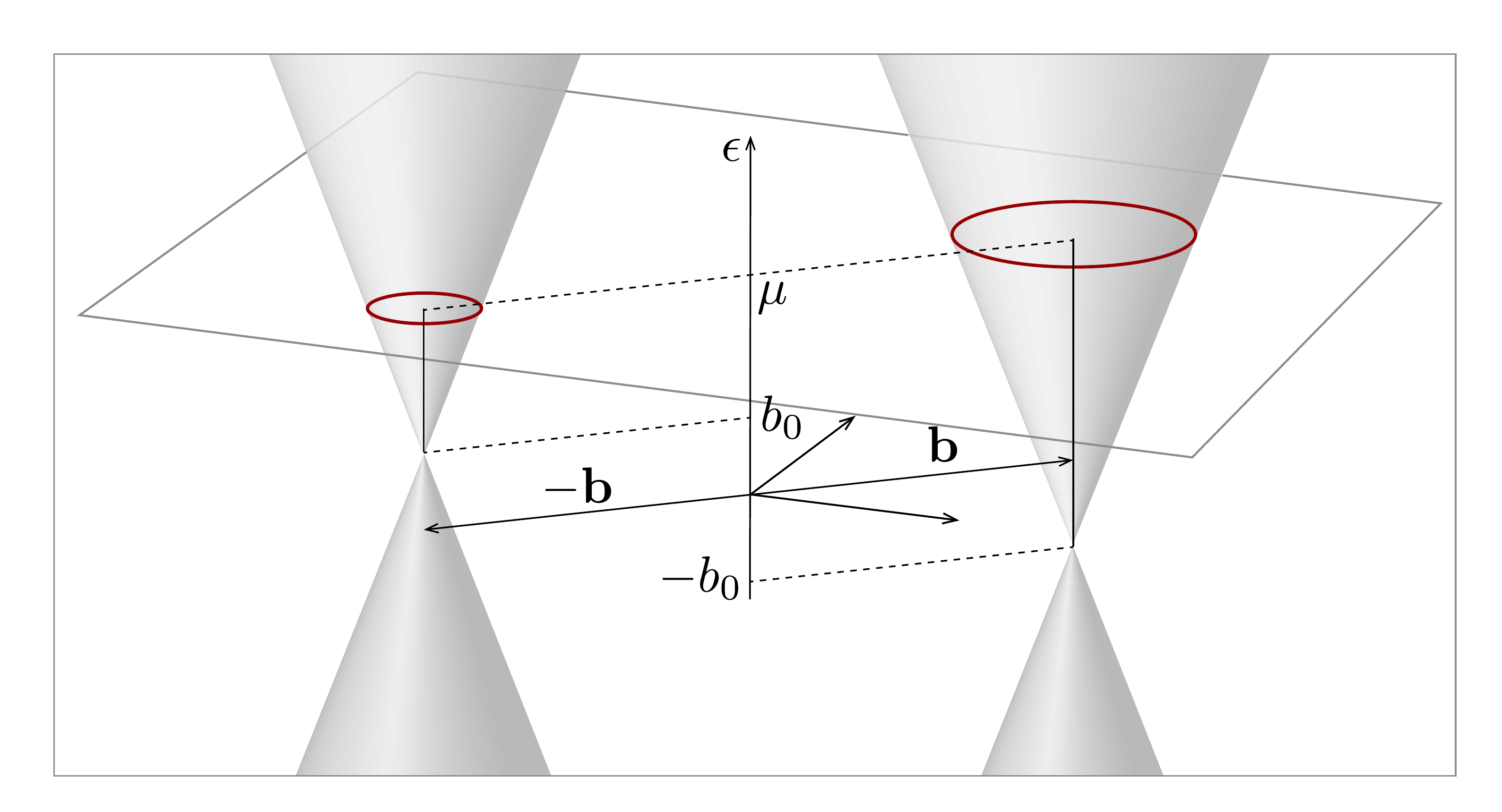}
  \caption{Two-dimensional schematic of the linearized dispersion relation characterizing a Weyl metal. Two (more generally, an even number of) Dirac cones, shifted relatively to each other both in momentum, $\mathbf{b}$, and/or energy, $b_0$, are embedded into a $3d$ Brillouin zone. 
The ensuing topological Fermi surfaces defined by the chemical potential $\mu$ 
(shown as red circles) are Fermi \emph{spheres} in $3d$ reality.}
  \label{fig1}
\end{figure}

All these features are shown by the Weyl semimetal, a system
that has been experimentally realized~\cite{Bernevig:2015,Xu:2015,Yang:2015,Lv:2015,Xu:07082015,Lv:2015a}
and is attracting a lot of attention~\cite{Weng:2015,Huang:2015,Burkov:2015,Hosur2013Recent,Souma:2015,Huang:2015a,Xiong:2015,Zhang:2015exp,
Yang:2015exp,Zhen:2015,Lu:2015,Klier:2015,Baum:2015,Sbierski:2014,Pesin:2015}. 
The topological centers of the Weyl semimetal are two Weyl points in its three
dimensional Brillouin zone (cf. Fig.~\ref{fig1}.) These points are monopoles of geometric phase and
therefore cannot be separately  gapped out. 
Much of the physics of the Weyl system is related to the exchange of charge between
individual nodes, even if the nodes are not connected by direct scattering. From a
condensed matter perspective, this phenomenon is understood as a manifestation of
spectral flow, i.e. occupation number altering rearrangements of the spectrum  under
the influence of, e.g., external magnetic and electric fields. From a perspective focusing on the
low energy effective Weyl Hamiltonians of the individual nodes, the same phenomenon is understood as a consequence of
the axial anomaly. Two prominent manifestations of the anomaly and the parity
non-invariance of individual Weyl nodes are the so-called chiral magnetic effect
(CME)~\cite{Alekseev:1998, Fukushima:2008} and the anomalous Hall effect (AHE)~\cite{Nagaosa:2010} respectively.

Much of our understanding of the anomaly in the Weyl semimetal, and of the ensuing
physical phenomena has been developed for the idealized clean system. On the other
hand, it is evident that impurity scattering must have profound influence on the
structure of the Dirac spectrum in the vicinity of the nodes. One may argue that this
cannot compromise the topological charge carried by the nodes, and hence will not
affect physical observables anchored in topology. However, short range correlated
impurity potentials have the capacity to scatter charge carriers \emph{between} the
nodes, and such type of scattering does spoil the topological protection. The joint influence of intra- and inter-node scattering characterized by rates $\tau$ and $\tau_\mathrm{n}\gg \tau$\footnote{In experiment~\cite{Xiong:2015} $\tau_n/\tau\sim 50$.}, respectively, was studied in pioneering work by Burkov~\cite{Burkov2014Chiral} and Son and Spivak~\cite{Son:2013}. 
Focusing on regimes of large chemical potential $\mu \gg \tau^{-1}$, they studied signatures of the CME in the diffusive longitudinal magnetoconductivity, $\sigma$, and obtained a contribution of topological origin, $\delta \sigma \sim \tau_n B^2$, where $B$ is an external magnetic field colinear with the direction of the field gradient and the current flow. This is a remarkable result which shows that transport phenomena of topological origin remain visible deep in the diffusive regime and, in fact, for any finite inter-node scattering rate. On the other hand, the parametric dependence of the correction raises the question what happens at large magnetic fields and/or in the limit of vanishing inter-node scattering rate $\tau_\mathrm{n}\to \infty$. 

An answer has been formulated in Ref.~\onlinecite{Parameswaran:2015} from the complementary perspective of the nearly clean limit, in which $\tau_n^{-1}$ is negligibly small, but $\tau^{-1}$ may remain finite. In this case, the presence of a magnetic field causes the formation of Landau levels (LL), and the opposite chirality of the lowest lying LL at the two nodes implies the onset of a drift current. The magnitude of the current is proportional to the LL degeneracy, and this translates to an effectively ballistic conductivity $\sigma\stackrel{\tau_n\to \infty}\sim L B$, where $L$ is the system extension in field direction. 

The two results  $\delta \sigma \sim \tau_n B^2$ and  $\sigma\sim L B$ do not match
trivially and explaining how they can be reconciled with each other will be one of
the  objectives of this paper. We will find that the matching problem, and in fact
various other unconventional transport signatures of the system, can be explained in
terms of an effective drift-diffusion crossover dynamics, which in turn originates in
a competition of impurity backscattering and topological current flow. What makes
this phenomenon unusual, and to the best of our knowledge unique to the Weyl system,
is that diffusion at \emph{short} length scales (yet larger than the elastic
scattering mean free path) crosses over to effectively ballistic dynamics at
\emph{large} scales. (I.e. the situation is opposite to that in conventional
scattering environments where ballistic motion at short scales crosses over into
diffusion at larger scales.) 

The formation of a drift diffusion regime provides the key to the solution of the above matching problem, and leads to a number of rather unconventional transport phenomena, which have not been discussed so far. In this paper, we will understand this physics within the context of the global phase diagram of the strongly disordered Weyl system. This picture will in turn be obtained from a microscopically derived field theory, which in many ways resembles that of a three-dimensional disordered Anderson metal. The notable difference lies in the presence of two types of topological terms, which reflect the topological charge of the Weyl nodes of the system. These terms support, respectively, the AHE and the CME. They show a high degree of robustness to impurity scattering, and in confined geometries can even overpower the effect of Anderson localization. The observable consequence are anomalies in long range transport coefficients, some of which have already been addressed within diagrammatic perturbation theory. In this paper, we will explore what happens in regimes beyond the reach of perturbation theory, and establish novel manifestations of topology in transport. Of these, the most remarkable is the above diffusion/drift crossover, which we will discuss in detail.

The rest of the paper is organized as follows. In view of its  volume, a qualitative
summary of our main findings is given in the introductory section~\ref{sec:MR}. In
section~\ref{sec:FTA} we introduce the field theoretical approach and derive the low
energy action of the system. We have tried to keep the discussion as non-technical as
possible, but self-contained; several intermediate steps are relegated to appendices.
We will also discuss the behavior of the theory under renormalization, which provides
us with a firm basis to establish its phase diagram. In section~\ref{sec:GFT}, we
couple the theory to external fields, and source fields required to compute
observables. We will also discuss the variational equations derived from the theory,
which on the one hand contain rather pronounced ‘geometric structure’ reflecting the
interplay of topology and the chiral anomaly, and on the other hand provide the key
to our subsequent description of transport. Anticipating the formation of novel types
of non-equilibrium transport, we extend the theory to a real time Keldysh formulation
in section~\ref{sec:Keldysh_FT}. This will be the basis for the discussion of
transport phenomena, where the focus will be on  the conductance and its noise
characteristics in the presence of external fields. We conclude in
section~\ref{sec:Conclusions}.

\section{Summary of main results}
\label{sec:MR}

In this paper, we will consider Weyl (semi)metals at length scales larger than the elastic mean free path $l$, i.e in regimes governed by multiple impurity scattering. Let us first consider the relatively simple situation in which the two Weyl nodes are strongly coupled, $\tau_n\to 0$. The physics of this limit is best understood by conceptualizing~\cite{Burkov2011Weyl} a three dimensional Weyl system as a stack of two-dimensional topological insulators. Each layer is governed by an anomalous quantum Hall effect (QHE), so the three-dimensional extension resembles a layered quantum Hall system similar to that discussed in Ref.~\onlinecite{Chalker:19915} in connection with the standard integer QHE. In that work it was shown that the three dimensional extension of the two dimensional quantum Hall insulator supports a metallic phase, and this turns out to be the phase relevant to  the three dimensional Weyl metal. The quantum Hall metal differs from a conventional metal by a non-vanishing Hall conductivity, which two-loop renormalization group analysis~\cite{Wang1997Localization} has shown to remain un-renormalized by disorder. (In this regard, the layered system differs from the two-dimensional QHE, for which the Hall conductivity renormalizes to integer values.) In the present context, the Hall conductivity does not require an external magnetic field, it is, rather, proportional to the separation $b$ of the Weyl nodes in momentum space, the AHE. Below, we will establish the above correspondence  by mapping the Weyl system onto the low energy effective field theory  of the layered QHE. The stability of both the longitudinal and the transverse conductivity with regard to disorder then follows from the earlier renormalization group study~\cite{Wang1997Localization}.

\begin{figure}[t]
  \centering
\includegraphics[width=7.5cm]{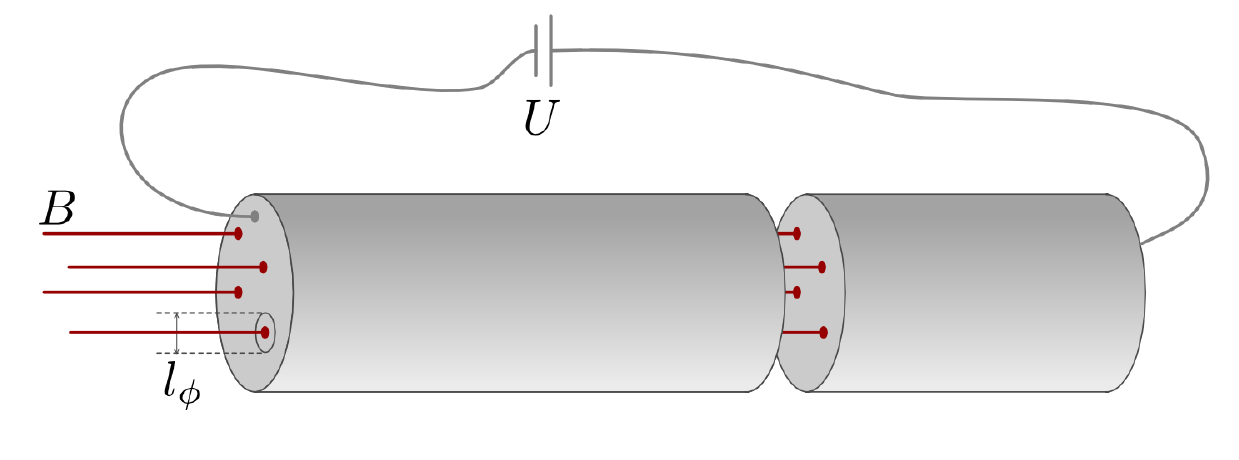}
  \caption{The application of a magnetic field ${\bf B}$ to a quasi-one-dimensional Weyl wire
creates left \& right moving channels propagating parallel to ${\bf B}$ which are
immune to \emph{intra}-node disorder scattering. Each channel has a characteristic
geometric cross section $l_B \times l_B$, so that their number equals $N_\phi =
\mathcal{A}/ 2\pi l_B^{2}= \Phi/\Phi_0$ --- the number of magnetic flux quanta piercing the
wire through the cross section ${\cal A}$, see Sec.~\ref{sec:Usadel_K} for more
details.}
  \label{fig:1DCMESchematic}
\end{figure}

The situation becomes  more interesting, once the condition of rigid mode
locking $\tau_n\to 0$ is relaxed. Dimensional analysis tells us that in this regime
three different length scales $l_n,l_m,l_\ast$ need to be discriminated. To see this,
consider the setup shown schematically in Fig.~\ref{fig:1DCMESchematic}: a Weyl metal
is subjected to a magnetic field of strength $B$. We are interested in its conduction
properties, and in particular the conductance in the direction of the field. For a
conventional metal, the conductance would be Ohmic, $g \sim  {\cal A} \nu D/L$, where $\nu$ is
the bulk density of states at the Fermi surface, $D$ is the diffusion constant, 
${\cal A}$ is the cross section and $L$ is the
length of the system. However, in a (clean) Weyl metal we have a different situation.
As pointed out in Ref.~\onlinecite{Parameswaran:2015}, the joint application of a magnetic and an electric
field leads to a particular manifestation of the axial anomaly, viz. the formation of
$N_\phi$ ideally conducting quantum channels, $g\sim N_\phi \times 1$, where $N\phi
\sim B \mathcal{A}$ is the number of flux quanta through the system. The equality of
these two expressions defines a length scale $l_m \sim \nu D/B$ suspected to
separate a diffusion dominated transport regime at short lengths from ballistic drift
dominated transport at large lengths.

\begin{figure}[b]
\includegraphics[width=8.0cm]{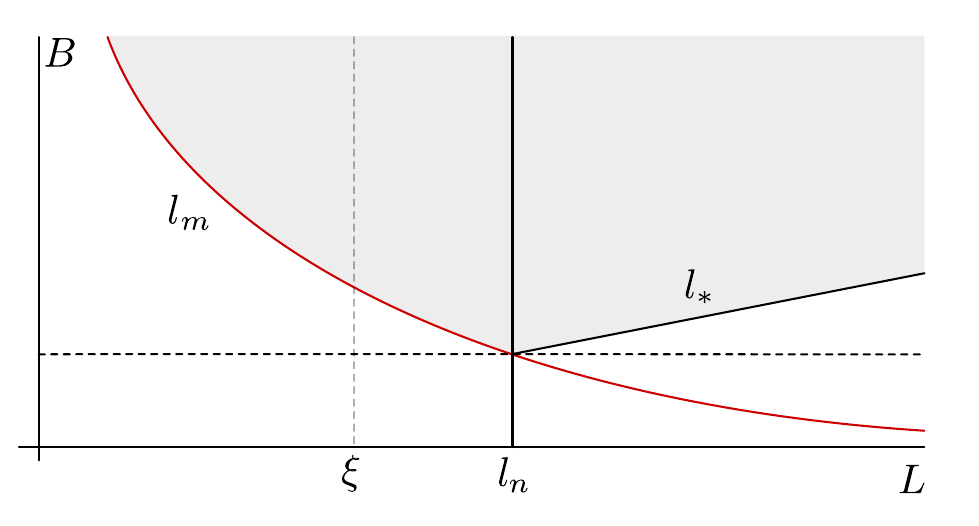}
\caption{Phase diagram of a quasi one dimensional Weyl quantum wire. The effectively realized transport regime depends on the length of the system relative to the nodal scattering length, $l_n$, at which the nodes get effectively coupled, and on the drift-diffusion crossover length $l_m\propto B^{-1}$. In the gray shaded area transport is effectively ballistic. We argue that in this regime the system will not show Anderson localization even if the localization length $\xi$ is smaller than $l_n$.} 
\label{fig:PhaseQuasiOne}
\end{figure}

While the drift dominated regime is robust with regard to \emph{intra}-node
scattering, it responds sensitively to inter-node scattering, and it ceases to exist
at length scales where the nodes get strongly coupled. Naively, one may suspect this
scale to be given by $l_n \sim (D\tau_n)^{1/2}$, i.e. the length scale accessible to
diffusive transport before inter-node scattering kicks in. However, the actual answer
turns out to be a little different. If $l_m \gg l_n$, then the nodes hybridize before the
ballistic transport can become effective. In this case, the presence of the latter
merely manifests itself in a small correction $\sim\tau_n B^2$ to the Drude conductance. 
However, if the field is strong enough such that $l_m \ll l_n$, then a drift regime in which transport is governed by the $N_\phi$ channels mentioned above exists. In this case, the question we need to ask turns out
to be when $N_\phi$ equals the number of states effectively hybridized by inter-node
scattering. For a system of length $L$, 
the latter equals $1/(\tau_n \delta_L)$ --- here $\delta_L = (\nu\mathcal A L)^{-1}$ is the mean level spacing
in the volume $\mathcal A L$, --- 
and the equality $1/(\tau_n \delta_L)  \sim N_\phi$ defines a crossover scale
$l_\ast \sim  \tau_n B/\nu \sim l_n^2/l_m$. For system of length $L\gtrsim l_\ast$ a re-entrance from
ballistic to diffusive transport takes place. We thus arrive at the conclusion that
for $l_m \ll l_n$ the system supports an extended regime of drift transport, indicated as
a gray shaded area in Fig.~\ref{fig:PhaseQuasiOne}. Below we will establish that in
the drift regime the conductance is length and disorder independent, \emph{and}
noiseless, i.e. it satisfies the criteria of genuine ballistic transport. In fact, we
will argue that in confined geometries where the system resembles a quasi-one
dimensional wire as in Fig.~\ref{fig:1DCMESchematic} the drift even overpowers
Anderson localization in the (perhaps fictitious) case where the localization length
$\xi\sim \nu D \mathcal{A}$ is smaller than $l_n$.

While the length scale $l_\ast$ can be defined on dimensional grounds, its identification as a crossover length requires justification. Below, we will see that for any finite $\tau_n>0$ the effective field theory describing the disordered system splits into two sectors, one for each node. The effective actions describing these nodes will be identified as variants of a Chern-Simons action, containing an ordinary diffusion term, and a Chern-Simons term which contains the information on the parity breaking part of the nodal physics. In the presence of a magnetic field, the Chern-Simons term gives rise to contributions to the action which contain only one derivative -- a hallmark of ballistic transport -- and overpower the two-derivative diffusion term at large length scales. This is the technical reason behind the formation of the above crossover regimes. In particular, the length scale $l_\ast$ will be that where the drift term and a term coupling the two nodal sectors by scattering balance each other. At larger scales, the nodal fields get effectively locked, and the respective Chern-Simons actions cancel out due to their opposite sign, which in turn is a signature of the opposite nodal parities.

This concludes our preliminary sketch of the results discussed in this paper. The
remainder of the text can be read in different ways: readers primarily interested in
results may directly turn to sections~\ref{sec:Renormalization} and
~\ref{sec:Transport}, where the long distance fluctuation behavior and the transport
physics of the system are discussed, respectively. Readers wishing to see the theory behind this, find its construction (sections~\ref{sec:FTA} and \ref{sec:GFT}) and analysis (section~\ref{sec:Keldysh_FT}) in the next sections. For the sake of
transparency, a number of technical construction steps are relegated to appendices.

\section{Effective field theory}
\label{sec:FTA}

\subsection{Model and disorder averaging}
\label{sec:Model}

The low-energy spectrum of the Weyl semimetal contains (at least) two topologically
protected nodes in the Brillouin zone. Introducing a  momentum cutoff $\Lambda\ll a_0$  smaller than the microscopic lattice spacing $a_0$ such that for momenta $|\bk|<\Lambda$ off the nodal centers the spectrum can be linearized, we describe a paradigmatic bi-nodal Weyl system by the Hamiltonian
\begin{equation}
\label{eq:H}
 	\hat H = v \sla{\hat{k}}\, \sn_3 +(v\sla{b} + b_0)+  V (\bx) = \hat H_0 + V (\bx),
 \end{equation} 
where the two nodes are split by the vector $2\bb\equiv 2b\mathbf{e}_3$ along the
 $z$-direction in momentum space, and by an offset $2b_0$ in  energy
 (Fig.~\ref{fig1}). We interchangeably use $(1,2,3)$ and $(x,y,z)$ to label coordinate directions, and the standard relativistic notation $\sla {\hat k}\equiv
 \hat\bk\cdot \bsigma$, where $\bsigma = (\sigma_1, \sigma_2, \sigma_3)$ denotes a
vector of Pauli matrices, $\hat \bk = -i\partial_{\bf r}$ is the momentum operator.
The non-universal velocity $v$ is fixed by the band structure, and the Pauli matrix
$\sn_3$ acts in a two-component space defined by the nodal structure.

We model the presence of disorder by a scalar Gaussian distributed potential $V (\bx)$ with a
variance $\gamma_0$,
\begin{equation}
\langle V (\bx) \rangle = 0, \quad \langle V (\bx) V (\bx') \rangle = \frac{\gamma_0}{2} f(|\bx - \bx'|/l_0),
\end{equation}
where the correlation function $f$ is normalized to unity, $\int dx f(|\bx|/l_0) =
1$, and the correlation radius $l_0$ much larger than the lattice constant $a_0$. Throughout, we use the abbreviation $\int dx\equiv \int d^3x$ for three dimensional volume integrals.

Our goal is to derive an effective theory describing the disorder averaged system. To
this end,  we  introduce the replicated partition function
\begin{equation}
\label{eq:ZR}
Z^R=\int D(\bar \psi,\psi)\,\exp(-S[\bar \psi,\psi])
\end{equation}
representing  $r=1,\dots,R$ identical Weyl fermions of energy $\epsilon$ in terms of the Gaussian action
\begin{equation}
\label{eq:S}
	S[\bar \psi,\psi]=-i\int dx \,\bar \psi(\epsilon+i \delta \tau_3 - \hat H)\psi.
\end{equation}
Here $\delta \to 0^+$, $\psi=\{\psi_{s,i,n}^r(\bx)\}$ is an $8R$-component vector of
Grassmann  fields, the index $n=1,2$ labels two nodes,  $i=1,2$ denotes the
components of a Weyl spinor, and $s=\pm$ is an index discriminating between the
advanced and retarded Green functions we need to access transport properities, i.e. $
(\tau_3)_{ss'}=s \delta_{ss'}$. 
From the functional~(\ref{eq:ZR}), expectation values of of currents may be computed by substitution $\partial_{\bf r} \to \partial_{\bf r} - i {\bf a}$, where the `vector potential' $\mathbf{a}$ is a suitably constructed source variable. In a manner discussed in more detail in later sections, differentiation w.r.t. $\mathbf{a}$ then yields the required quantities. However, to keep the notation simple, we suppress the source dependence of the functional for the time being. (For completeness, we mention that in the absence of sources, the integration over $\psi$ yields $Z^R=\det(G^+G^-)^R$, where $\hat G^\pm=(\epsilon\pm i\delta -\hat H)^{-1}$ is the resolvent operator. Replica analytic continuation $R\to 0$ then leads to a unit normalized $\lim_{R\to 0}Z^R=1$.) We finally mention that the pre-averaged Gaussian functional affords alternative representations, viz. as a supersymmetric, or a Keldysh functional. In Sec.~\ref{sec:Keldysh_FT}, we will discuss the straightforward adaption from replicas to Keldysh and employ the latter variant for the computation of nonequilibrium transport characteristics.

The Gaussian average over disorder generates the 4-fermion `interaction' vertex 
\begin{equation}
\label{eq:S_dis}
S_{\rm dis} = \frac{\gamma_0}{2}\sum_{n=1,2}\int dx\, (\bar\psi_n\psi_n)^2 + 
{\gamma_n}\int dx\, (\bar\psi_1\psi_1)(\bar\psi_2\psi_2),
\end{equation}
where two terms describe the \emph{intra} and \emph{inter} node scattering respectively. In cases where the correlation length is long, $l_0 \gg b^{-1}$, the  coupling strength strength of the latter
\begin{equation}
\label{eq:gamma_n}
\gamma_n = \int dx f(|\bx|/l_0)e^{i {\bf b}\bx}
\end{equation} 
is exponentially suppressed  relative to the inter-node amplitude $\gamma_i\ll
\gamma_0$. This limit, which is typically realized in current experiments,  will be
assumed throughout the paper.

We start by considering the completely decoupled limit, $\gamma_n\to 0$, in which the
theory separates into two disconnected sectors, $n=1,2$, each describing an isolated
Weyl node. The effect of  inter-node scattering will be added at a later stage in
Sec.~\ref{eq:real_time}. Focusing on node $n=1$, and suppressing the index $n$ for
notational simplicity we
decouple~\cite{Altland1999Supersymmetry,Efetov1997Sypersymmetry} the quartic vertex
describing  inter-node scattering by a matrix field
$B(\bx)=\{B^{rr'}_{ss',ii'}(\bx)\}$ describing the phase coherent propagation of
particle-hole amplitudes $\psi^r_{s,i}(\bx)\bar
\psi_{s',i'}^{r'}(\bx)$ in the disordered background. Gaussian integration over $\psi$ then yields the averaged partition
sum $\langle Z^R \rangle = \int {\cal D} B \exp(-{\cal S}[B])$ with
the effective action
\begin{equation}
\label{eq:S_B}
{\cal S}[B]=\frac{1}{2\gamma_0}\int dx \,\mathrm{tr}B^2-\ln \mathrm{det}\, \hat G[B].
\end{equation}
Here $\hat G[B]=(\epsilon + i\delta \tau_3 -\hat H_0-  B)^{-1}$ is the $B$-dependent Green's function 
and $\hat H_0$ stands for the clean Hamiltonian, see Eq.~(\ref{eq:H}).

We proceed by subjecting the action ${\cal S}[B]$ to a saddle point analysis. (For
remarks on the validity of the latter, see below.)  Variation of the
action~(\ref{eq:S_B}) w.r.t. $B$ yields the mean field equation
\begin{equation}
\label{eq:SCBA}
\bar B\stackrel{!}{=}\gamma_0\,{\rm tr}_{\rm }\,\hat G({\bf x,x};[\bar B]),
\end{equation} 
which can be solved in terms of the diagonal ansatz, $\bar B=-i \kappa \tau_3$.
The mean field equation~(\ref{eq:SCBA}) is known to be equivalent to the self consistent Born approximation (SCBA)
and its solution $\bar B$ plays the role of an impurity self energy within the `non-crossing approximation'. Evaluating the equation in the specific limit of zero energy, i.e. the Weyl semimetal limit, one obtains~\cite{Fradkin1986Critical}
\begin{eqnarray}
	-i{\kappa} &=& \gamma_0 \int_{|{\bf p}|<\Lambda} \frac{d^3p}{(2\pi)^3}\,\mathrm{tr}\left(\frac{1}{i\kappa-v\sla p}\right)\nonumber \\ 
	&=& -\frac{i\gamma_0 \kappa\Lambda}{(\pi v)^2}\left(1-\frac{\kappa}{v\Lambda}\arctan\left(\frac{\Lambda v}{\kappa}\right)\right).
\end{eqnarray}
This equation has a non-zero solution only for disorder strength exceeding the
critical one $\gamma_* = (\pi v)^2/\Lambda$. For $\gamma_0$ slightly larger than
$\gamma_*$ the self-energy reads $\kappa = (2/\pi) v\Lambda
(1-{\gamma^\ast}/{\gamma_0})>0$. 
The presence of a critical disorder strength $\gamma_*$  is consistent
with renormalization group (RG) studies in $d=2+\epsilon$ dimensions~\cite{Syzranov2014Localisation,Roy:2014} at $\epsilon=1$. At one-loop order the RG equation for the running coupling $\gamma$ has the form
\begin{equation}
\frac{d\gamma}{d\ln(1/\Lambda')} = -\gamma  + \frac{\gamma^2}{\gamma_*},  
\end{equation}
which shows that as one lowers a running momentum cut-off $\Lambda'$ the effective
disorder strength decreases or increases depending on whether the bare value
$\gamma_0$ is greater or smaller than the critical one. If $\gamma_0 < \gamma_*$ the
$\beta$~-~function has a stable IF fixed point $\gamma=0$, while in the case
$\gamma_0>\gamma_*$ the RG equation predicts flow towards a strong disorder regime.
The perturbative RG treatment of this flow ceases to be valid at a cutoff equal to the inverse of an effective `mean
free path'  $\Lambda^{-1}\equiv l$ defined by the condition that at the cutoff scale the kinetic part of the fermion action is
comparable with the disorder scattering vertex. Inspection of the fermionic action shows that this leads to the implicit condition $l \sim
\gamma(l)/v^2$. For $\Lambda \sim l^{-1}$ there are
no large parameters in the problem left, which implies that $\gamma(l) \sim
\gamma_\ast$ exceeds the critical coupling strength only by numerical factors (as can be checked by direct integration of the RG equation.)

Throughout, we will be interested in the physics at large length scales $\gg l$,
which is governed by multiple scattering and described by soft fluctuations around
the non-vanishing mean field. Following to the logics above, we commence the analysis
of the field theoretical action at an effective renormalized disorder amplitude $\sim
\gamma(l)>\gamma_\ast$. Even so, the field theory we derive at the Weyl semimetal
point will turn out to be governed by strong fluctuations at the shortest scales
$\sim l$, and flow towards weaker fluctuations only at larger scales. This means that
at the bare level the derivation of the theory is only poorly controlled. By
contrast, for chemical potentials $\mu \gg \gamma_\ast$ away from the Weyl node (i.e.
in the Weyl metal regime) one obtains a non-vanishing self-energy $\kappa = \gamma\pi
\nu$ at any disorder strength where $\nu = \epsilon^2/2\pi^2 v^3$ is the clean
density of states of the $3d$ Dirac Hamiltonian. In this regime, the derivation of
the theory is well controlled. Since we do not expect phase transitions upon lowering
the chemical potential (for fixed disorder), the stability of the large $\mu$ analysis corroborates
the validity of its $\mu=0$ limit.

\subsection{Effective action}

The non-zero mean field solution $\bar B$ breaks the original 'replica rotation
symmetry' of the action~(\ref{eq:S}), which is invariant under global unitary
transformation $\psi \to U\psi$, where $U$ is the spatially constant matrix from the
group $G = U(2R)$ acting in the direct product of the replica and advanced retarded
space. Besides, the saddle point solution $\bar B$ is not unique. The full manifold
of saddle points is parametrized by $B = i\kappa T \tau_3 T^{-1}$ with $T$ being the
element of the group $G$. We see, however, that among all possible $T$'s the subgroup
of matrices $k \in H \equiv U(R) \times U(R)$ commutative with $\tau_3$ does not
affect the SCBA result. Thus one concludes that all non-empty fluctuations $T \in
G/H$ form a manifold of Goldstone modes, later to be identified as diffusively propagating soft modes.

In the next subsection we derive the low energy field theory as the (regularized)
expansion of the action~(\ref{eq:S_B}) in terms of generators $T^{-1}\partial_i
T\equiv A_i$ of Goldstone mode fluctuations $T(\bx)$. The resulting theory 
contains a conventional two-gradient term  (known as a diffusion term in the present
context), plus additional contributions of topological origin. For the sake of reference we state our result for the effective action of the system here, before its derivation is discussed in the next section. As long as we ignore inter-node scattering, $\gamma_n=0$, the action splits into a sum $S[A_1,A_2]=S_1[A_1]+S_2[A_2]$ of two independent nodal fields $A_n$. The nodal actions are in turn given by the sum of three pieces,
\begin{align}
\label{eq:GlobalActionStructure}
 S_n=S_{\mathrm{d}}+S_{\mathrm{top}}[A]+(-)^n S_{\mathrm{CS}},  
\end{align}
of which only the third shows dependence on the nodal index via a parity sign change. The first term in~\eqref{eq:GlobalActionStructure} is given by 
\begin{eqnarray}
\label{eq:Q_action_diff}
S_\mathrm{d} [Q] &=&
\frac{\sigma_{xx}}{8}\int dx\,\mathrm{tr}(\partial Q^2), \nonumber\\
&&\sigma_{xx} = \frac{\epsilon^2 + 3 \kappa^2}{6\pi\kappa v}.
\end{eqnarray}
It was first derived in
Ref.~\onlinecite{Fradkin1986Critical} and describes the diffusive dynamics of low energy excitations, in terms of a ‘stiffness’ determined by the longitudinal SCBA conductivity $\sigma_{xx}$ of the system. 

The second term
\begin{eqnarray}
\label{eq:Q_action_top}
S_\mathrm{top}[Q] &=&\frac{\sigma_{xy}}{8}
 \epsilon^{3ij}\int dx\,\mathrm{tr}
(Q\partial_i Q \partial_j Q),\nonumber \\
&&\sigma_{xy} = b/2\pi,
\end{eqnarray}
is of topological origin~\footnote{Strictly speaking, the three dimensional $S_{ \mathrm{top} }$ is not a topological term, only its two-dimensional projection, i.e. a term which would be obtained from $S_{\mathrm{top}}$ by ignoring the third coordinate and the integration over it, is a genuine $\theta$-term. However by  abuse of language we continue to call $S_{\mathrm{top}}$ ‘topological’.} and known from
the study of multilayer electron systems in the integer quantum Hall
regime~\cite{Wang1997Localization}. The appearance of this action in the present
framework can be understood from the fact that the Weyl system can be realized in
terms of stacked and coupled  2d quantum anomalous Hall insulators along
$z$-direction~\cite{Burkov2011Weyl}. The above action then described the ensuing
layered quantum Hall system.  Notice that the coupling constant $\sigma_{xy}$, which we will later relate to the anomalous Hall conductivity of the system, depends on the nodal splitting, $b=\mathbf{b}$, even if the nodes remain uncoupled by the Hamiltonian. As we will see, this is one of the ramifications of the anomaly in the system. We also note that the coefficients $\sigma_{\mu \nu}$ represent the contribution of a single node to the conductivity tensor of the system. In a system comprising two (or more generally $2n$) nodes, the full response coefficients are twice as large ($2n$ times as large.)

Finally, the third term
\begin{align}
    \label{eq:A_action_CS}
    &S_{\mathrm{CS}}[A]=S^I_{\mathrm{CS}}[A]+S^{II}_{\mathrm{CS}}[A],\\
    &\;S^I_{\mathrm{CS}}[A]=-\frac{i\epsilon^{ijk}}{8\pi}\sum_{s=\pm} s\int
    dx\,\mathrm{tr}
    (A_i P^s
    \partial_j A_k P^s),\nonumber\crcr
    &\;S^{II}_{\mathrm{CS}}[A]=-\frac{i\epsilon^{ijk}}{12\pi}\sum_{s=\pm} s\int
    dx\,\mathrm{tr}
    (A_i P^s
    A_j P^s A_kP^s),\nonumber
\end{align}
does not afford a representation in terms of the $Q$-matrices. Instead, it is expressed in terms of the fields $A=T^{-1} \partial T$, and projectors
$P^\pm = (\mathds{1} + \tau_3)/2$ onto the retarded/advanced
sectors of the theory. Apart from the presence of these projectors, the action has
the typical ‘$AdA+\frac{2}{3}A^3$’ structure of a non-abelian Chern-Simons action.
Referring for more details to the discussion below, we note that the appearance of
this action is a consequence of the particular ‘triangular’ Feynman diagrams present
in the expansion of  (2+1) or (3+0) dimensional massive relativistic gauge field
theories.

Finally, in the presence of internode scattering, the action contains a term
\begin{equation}
\label{eq:Simp}
S_{\rm imp}[Q_1, Q_2] = - \frac{\pi \nu}{4\tau_n}\int dx \, {\rm tr} (Q_1 Q_2).
\end{equation}
coupling the nodal fields at a strength $1/\tau_n = \gamma_n \pi \nu$.

In the next section we  discuss the derivation of the above effective action. Readers
primarily interested in physical applications  may skip these sections and continue
reading directly from Sec xx, where we discuss how the action describes the physics of the system at large distance scales. After that, in
Sec.~\ref{sec:Keldysh_FT}, we proceed to discuss the coupling of the theory to external sources (the structure of which is conveniently
prescribed by gauge invariance) and its application to the computation of physical
observables, including the longitudinal or Hall conductivity, shot noise, and others.

\subsection{Regularization}

\label{sec:Reg}

The full action above derives from the  expansion of the 
determinant
\begin{equation}
\label{eq:S_Q}
S_0[Q]\equiv  - \mathrm{tr}\ln(\epsilon-v\sla k + i \kappa Q)
\end{equation} 
in soft $A$-fluctuations. (Due to $Q^2=\mathds{1}$, the Gaussian weight in the
action~\eqref{eq:S_B} reduces to an inessential constant.) A natural idea would be to
start by applying a similarity transformation to the Dirac operator and to consider
the action,
\begin{eqnarray}
S_0[Q]&{=}& - \mathrm{tr}\ln(T^{-1}(\epsilon-v\sla k + i \kappa
Q)T)\\
\label{eq:S_A}
&\stackrel{?}{=}&- \mathrm{tr} \ln(\epsilon-v\sla k+ i v \sla A + i \kappa
\tau_3)\equiv {\cal  S}_0[A],\nonumber
\end{eqnarray}
which then may be gradient expanded in powers of the `non-abelian gauge field' $A$.
However, because of the notorious UV divergences of the Dirac operator the
actions~(\ref{eq:S_Q}) and (\ref{eq:S_A}) are not equivalent. Thus our strategy must
be to first regularize the Dirac operator and only then proceed with the gradient
expansion. Following a construction  developed in  Ref.~\onlinecite{Altland2002Theories} in connection with the theory of disordered 2d $d$-wave superconductors, we
regularize the action~(\ref{eq:S_Q}) as
\begin{eqnarray}
\label{eq:S_Q_reg}
S[Q] &\equiv & S_0[Q]-S_\eta[Q] \\
& = &- \mathrm{tr}\ln(i \kappa Q + \epsilon-v\sla k) +
\mathrm{tr}\ln(i \eta Q -v\sla k ).\nonumber 
\end{eqnarray} 
Here $S_\eta$ differs from $S_0$ by a replacement $\kappa\to \eta$, where $\eta$ is
infinitesimal, and by setting $\epsilon = 0$. In the limit $\eta~\to~0^+$ the action
$S_\eta[Q]$ becomes $Q$-independent and gives an inessential constant. However, for
large momenta $|{\bf k}|\gg \kappa/v$, the contributions from the two actions
$S_0[Q]$ and $S_\eta[Q]$ cancel against each other and the full action $S[Q]$ becomes
UV finite.

One may now safely proceed with the similarity transformation applied to both terms
of the regularized action to obtain
\begin{align}
\label{eq:action_reg}
&\mathcal{S}[A] \equiv {\cal S}_0[A]-{\cal S}_\eta[A] \\
&= - \mathrm{tr}\ln(i \kappa \tau_3 + \epsilon-v\sla k + i v\sla A) + 
\mathrm{tr}\ln(i \eta \tau_3 - v\sla k + i v\sla A). \nonumber
\end{align} 
The action $\mathcal{S}[A]$ is structurally similar\footnote{In
Ref.~\onlinecite{Redlich1984Parity}, the bare action $\cal S_\eta[A]$ of massless
fermions moving in the background vector potential $A$ was regularized by the
subtraction of a bosonic action ${\cal S}_{0}[A]$ containing a formally infinite mass
$(M \equiv -i\kappa) \to \infty$. This choice of the regularization scheme preserves
the gauge invariance of the theory but breaks the so-called parity symmetry $P$.
Defining Clifford $\gamma$-matrices in $d=2+1$ dimensions as $\gamma_0=\sigma_3$ and
$\gamma_{1,2} = \sigma_{1,2}$, the latter is defined as $\psi(x,y,t) \to \gamma_2
\psi(x, -y,t)$ and $\bar\psi(x,y,t) \to - \bar \psi(x, -y,t) \gamma_2 $. Under the
$P$--symmetry the vector potential is transformed as $A_{y} \to - A_y$ while
$A_{x,t}$ do not change. The action $\cal S_\eta[A]$ is $P$--invariant in the limit
$\eta \to 0^+$ but the finite mass $M$ in ${\cal S}_{0}[A]$  explicitly breaks it.
The key message here is that one may regularize maintaining gauge invariance
\emph{or} parity invariance, but not both. The breaking of parity symmetry by UV
quantum fluctuations goes by the name ‘parity anomaly’.} to the (regularized) action
of a 3d Dirac operator in a background nonabelian gauge field. This action was
analyzed in the  classic Ref.~\onlinecite{Redlich1984Parity}, with the main result
that the ensuing effective action for the gauge field $A$ is of Chern-Simons type. In
the following, we will demonstrate that a Chern-Simons term is indeed present, next
to two others following from the particular form of our gradient fields.

All these terms follow from the expansion of the now regularized fermion determinant
in powers of $A$. To formulate this expansion, we first define the SCBA Green
function $G_p\equiv (\epsilon-v(\sla p+\sla b)+  i \kappa
\tau_3)^{-1}$ describing the propagation of excitations of momentum $\mathbf{p}$ relative to the nodal momentum $\mathbf{b}$, damped by the impurity self energy $i \kappa$. Straightforward matrix algebra yields
\begin{align}
\label{eq:Green_SCBA}
    G_p= \frac{\epsilon + i \kappa\tau_3 + \sla{p} +\sla{b}}
{(\epsilon +  i\kappa\tau_3)^2-({\bf p} + {\bf b})^2} = \sum_{s=\pm} G_p^s P^s,
\end{align}
where $G_p^\pm$ is the retarded/advanced Green function, $p\equiv (\epsilon,\bp)$,  and we temporarily set $v=1$
 to simplify the notation. Expansion of ${\cal S}_0[A]$ up to  third order (when
 probing structures at length scales $L$, higher orders in the expansion will be
 suppressed in powers of $l/L$) in $\sla{A}$ yields~\footnote{Our convention for the
 momentum integration measure is $dp \equiv d^3 p/(2\pi)^3$.}
\begin{eqnarray}
    {\cal S}_0[A]&=&{\cal S}^{(1)}[A]+{\cal S}^{(2)}[A]+{\cal S}^{(3)}[A] + \dots ,\\
    {\cal S}^{(1)}[A]&=&-i\,\mathrm{tr}(G\sla{A}),\nonumber\\
    {\cal S}^{(2)}[A]&=&-\frac{1}{2}\int dq dp \,\mathrm{tr}(G_{p+q} \sla{A}_qG_p \sla{A}_{-q}), \nonumber \\
    {\cal S}^{(3)}[A]&=&\frac{i}{3}\int dq_{1} dq_{2} dp \nonumber\\
&\times& \,\mathrm{tr}(G_{p+q_1+q_2} \sla{A}_{q_1}G_{p+q_2} \sla{A}_{q_2}G_p \sla{A}_{-q_1-q_2}). \nonumber
\end{eqnarray} 
(In the term $\mathcal{S}^{(1)}$ of first order in $A$ it is better not to prematurely switch to a momentum representation, cf. discussion in Sec.~\ref{sec:StopDer} below.)
In the following, we show how the three principal terms~\eqref{eq:Q_action_diff}, \eqref{eq:Q_action_top} and~\eqref{eq:A_action_CS} can be extracted from this formal expansion.

\subsection{Derivation of the diffusive action $S_\mathrm{d}$}
\label{sec:SdDer}

Inspection of the diffusive action~\eqref{eq:Q_action_diff} shows that this action contains two derivatives, and that no mixed derivatives $\partial_1 \partial_2$ are present. This indicates, that that part of the action is obtained from second order expansion in $A_i A_i$, with no further derivatives acting on the $A$'s. Terms of this structure are obtained  from the action ${\cal
S}^{(2)}[A]$ upon neglecting the slow $q$ momenta w.r.t. the fast $p$ momenta, i.e.
by setting $G_{p+q} \simeq G_q$. One then finds 
\begin{eqnarray}
\label{eq:S2_0}
{\cal S}^{(2)}_0[A] &=& -\frac{1}{2}\int dq dp \,\mathrm{tr}(G_{p} \sla{A}_{q} G_p \sla{A}_{-q}) \nonumber \\
&=& -\frac{1}{2}\sum_{i,ss'} f_{ss'} \int dx\, \mathrm{tr} (P^s A_{i}P^{s'} {A}_{i} ). 
\end{eqnarray}
Here,  the symbol $f_{ss'}$ represents an integral over the fast momentum variable $p$, which after the shift of integration  variables ${\bf p} \to {\bf p} - {\bf b}$ (which is save, because we deal with UV finite contributions) read as
\begin{eqnarray}
\label{eq:fssDef}
f_{ss'} &=& \int dp\, \mathrm{tr}\left[ (\epsilon_s + \sla{p})\sigma_i 
        (\epsilon_{s'} + \sla{p})\sigma_i \right] N_p^s N_p^{s'},
        \label{eq:f_ss}
\end{eqnarray}
where we have defined $N_p^s = ({\epsilon_s}^2 - p^2)^{-1}$ with $\epsilon_s =
\epsilon + i s \kappa$. The somewhat technical evaluation of this and a number of
similar integrals  is detailed in Appendix~\ref{sec:fast_momentum_integrals} and
leads to
\begin{eqnarray}
\label{eq:f_ss_res}
f_{ss} = 0, \qquad f_{s,-s} = (\epsilon^2 + 3\kappa^2)/6\pi \kappa. 
\end{eqnarray}
Given the definitions, $Q=T\tau_3 T^{-1}$ and $A_i = T^{-1}\partial_i T$, it is straightforward to check that
\begin{equation}
\label{eq:Id2}
\sum_{s=\pm} {\rm tr}\left( P^s A_i P^{-s} A_i\right) \!=\! -\frac{1}{4}{\rm tr}[\tau_3,A_i]^2 \!=\! 
-\frac{1}{4}{\rm tr}(\partial_i Q)^2.
\end{equation}
Substituting this expression into $S_0^{(2)}[A]$, we obtain the identity of the
latter with the diffusion term $S_d$ (\ref{eq:Q_action_diff}) with the coupling constant
$\sigma_{xx} =  f_{s,-s}$.

\subsection{Derivation of the topological action $S_\mathrm{top}$}
\label{sec:StopDer}

The topological term
$S_{\rm top}[Q]$, too, contains two derivatives. The two principal candidates for contributions to $S_\mathrm{top}$ therefore are  the terms of  ${\cal O}( A q)$ and ${\cal O}(A^2
q^0)$ in the expansion of the regularized action~(\ref{eq:action_reg}). As is known from the theory of the quantum Hall effect~\cite{Pruisken1984a, Levine1984Theory}, these two contributions have a distinct physical meaning:  the first couples to all states of below the Fermi energy. Its coupling constant, which is generally denoted by $\sigma_{xy}^{\rm II}$ is the second of two contributions to the celebrated Smr\v{c}ka-St\v{r}eda
formula~\cite{Smrska_Streda:1977} $\sigma_{xy}=\sigma_{xy}^{\rm I}+\sigma_{xy}^{\rm II}$ for the Hall conductivity. The second term describes the Hall response of states at the Fermi surface, and its coupling constant $\sigma_{xy}^{\rm
I}$ is the second contribution. In the following, we analyse these to terms separately. 

\emph{Fermi surface Hall response, $\sigma_{xy}^{I}$.} This contribution to $\sigma_{xy}$ derives from the 2nd order
action~(\ref{eq:S2_0}) which in full generality has the form
\begin{eqnarray}
\label{eq:S2_0_ij}
{\cal S}^{(2)}_0[A] &=& -\frac{1}{2}\sum_{ij,ss'} f^{ij}_{ss'} \int dx\, \mathrm{tr} (P^s A_{i}P^{s'} {A}_{j} ), 
\end{eqnarray}
where
\begin{equation}
\label{eq:fssprimeDef}
f_{ss'}^{ij} = \int dp\, \mathrm{tr}\left[ (\epsilon_s + \sla{p} + \sla{b})\sigma_i 
        (\epsilon_{s'} + \sla{p}+ \sla{b})\sigma_j \right] N_p^s N_p^{s'}.
\end{equation}
generalizes the expression $f_{ss'}$ defined in Eq.~\eqref{eq:fssDef}. The evaluation of the integral is detailed in Appendix~\ref{sec:fast_momentum_integrals} and leads to $f_{ss}^{i\not= j}=0$ and $f_{+,-}^{i\not=j}=f_{-,+}^{i\not=j}=\frac{2\kappa}{\pi^2 }\frac{b}{\Lambda}$. Using the identity  (see also Eq.~(\ref{eq:Id1})),
\begin{align} 
&\sum_{s =\pm } s \epsilon^{3 ij} \mathrm{tr}(P^s A_i P^{-s} A_j)  = \nonumber \\
&\epsilon^{3 ij} \mathrm{tr}(\tau_3 A_i A_j) =
-\frac{1}{4} \epsilon^{3ij} \mathrm{tr} (Q\partial_i Q \partial_j Q),
\end{align}
one realizes that the action~(\ref{eq:S2_0_ij}) is converted into the Pruisken term~(\ref{eq:Q_action_top})
where $\sigma_{xy}^I = f^{12}_{+,-}$. The result above implies the vanishing of this coefficient at infinite momentum cut-off $\Lambda$.

We thus conclude that the Hall coefficient does not receive contributions from Fermi-surface excitations; the Hall response is entirely due to the ‘thermodynamic’ contribution from states below the Fermi surface to be discussed next. 

\emph{Thermodynamic Hall response, $\sigma_{xy}^{II}$.} The Hall coefficient $\sigma_{xy}^{II}$ is related to the  1st order expansion of the action~(\ref{eq:action_reg}) in $\sla{A}$,
\begin{align}
    S^{II}_\mathrm{top}&\equiv -i \mathrm{tr}(G\sla{A})= i\int_{-\infty}^\epsilon d\epsilon'\,\mathrm{tr}(G(\epsilon')^2 \sla{A}),
\end{align}
where the SCBA Green function $G(\epsilon)$ is defined in~(\ref{eq:Green_SCBA}).
Following Pruisken~\cite{Pruisken1984a}, we have doubled the power of the Green
functions, $G(\epsilon)=-\int^\epsilon d \epsilon' \,G(\epsilon')^2$ to improve the
convergence of the ensuing momentum integrals. (This is the formal operation which
brings the states below the Fermi surface into play.) In evaluating the trace over
momenta, we need to take  the non-commutativity of $G_p$ and $A(\bx)$ into account.
We do so by using that the product of two operators $A$ and $B$ diagonal in
coordinates and momenta, respectively, can be semiclassically expanded
as~\footnote{Precisely speaking, the symbol $X(\bx,\bp) \equiv 2\int
d \by\,e^{2i \mathrm{y}\cdot \bp } X(\bx+\by,\bx-\by)$ refers to
the Wigner transform of an operator, and the expansion mentioned in the text is the
Moyal product expansion.} 
\begin{align}
\label{eq:Moyal}
(AB)(\bx,\bp)&= A(\bx) B(\bp) + \frac{i }{2} \partial_\bx
A(\bx) \partial_\bp B(\bp)+\dots,\cr 
(BA)(\bx,\bp)&= A(\bx) B(\bp) - \frac{i }{2}
\partial_\bx A(\bx) \partial_\bp B(\bp)+\dots,
\end{align}
where $A(\bx)$ and $B(\bp)$ are the
corresponding eigenvalues, respectively, and the ellipses denote terms of higher
order in Planck's constant (here set to unity.). Likewise, the trace of such operator
products affords the representation $\mathrm{tr}(AB)=\int dx dp\, (AB)(\bx,\bp)$.
Application of these to the product of momentum diagonal Green functions $G=G(\bp)$
and coordinate diagonal fields $A(\bx)$ appearing under our trace yields
\begin{align}
&(G  \sla{A}  G)(\bx,\bp) = G(\bp)\, \sla{A}(\bx)G(\bp) \\
&-\frac{i}{2}\partial_{p_i}G(\bp) \partial_{x_i}\sla{A}(\bx) G(\bp) 
+ \frac{i}{2} G(\bp) \partial_{x_i}\sla{A}(x)\partial_{p_i}G(\bp) + \dots \nonumber
\end{align}
With this, we obtain for the action
\begin{equation}
    S^{II}_\mathrm{top}[A]\simeq \frac{1}{2}\int_{-\infty}^\epsilon d\epsilon'\int dp \int dx\,\mathrm{tr}([G_{p}, \partial_{p_i}G_p] \partial_{x_i}\sla{A}),
\end{equation}
where the coordinate/momentum arguments are suppressed for clarity. 
For further discussion it is again useful to represent the Green's function in the form 
\begin{equation}
G^s_p = (\epsilon_s + \sla{p} + \sla{b})N_p^s,
\end{equation}
where as before, we have introduced $\epsilon_s = \epsilon + i\kappa s$ and 
$N_p = [\epsilon_s^2 - ({\bf p} + {\bf b})^2]^{-1}$, cf.~Eq~(\ref{eq:f_ss}). 
This gets us to the topological action
\begin{equation}
\label{eq:S_II_top}
    S^{II}_\mathrm{top}[A] = \frac{1}{2} \sum_{s,ij} \lambda^{ij}_s \int
    dx\,\mathrm{tr}( \partial_i A_j P^s),
\end{equation}
where coefficients $\lambda^{ij}_s$ comprise the integration over energy and momentum,
\begin{eqnarray}
\lambda^{ij}_s &=& \int_{-\infty}^\epsilon d\epsilon'\int dp \,
\mathrm{tr} \Bigl( [\, \sla{p} + \sla{b}, \sigma_i]\, \sigma_j \Bigr) (N_p^s)^2\nonumber \\
\label{eq:lambda}
&=& 4 i\epsilon^{ijk}\int_{-\infty}^\epsilon d\epsilon'\int dp \, (p_k + b_k) (N_p^s)^2. 
\end{eqnarray}
The momentum integral above is non-zero only if the integrand is an even function of
$p_1$ and $p_2$, and this enforces $k=3$. To simplify the action~(\ref{eq:S_II_top}) further we use the
relation $\epsilon^{3ij}\mathrm{tr}( \partial_i A_j) =0$, which can readily checked
by employing the identity $(\partial_i T^{-1}) T = - T^{-1} \partial_i T$ together
with the cyclic property of trace. Thus only the terms $\propto \tau_3$ stemming from
projector matrices $P^\pm = (\mathds{1} \pm \tau_3)/2$  will lead to a
non-vanishing result. With these remarks at hand we find
\begin{equation}
\label{eq:S_II_top_A}
    S^{II}_\mathrm{top}[A] = \frac{1}{2} \sigma_{xy}^{II} \sum_{ij} \epsilon^{3ij}\int dx\,\mathrm{str}(\tau_3\partial_iA_j),
\end{equation}
with an identification of the  the Hall conductivity as
\begin{equation}
\label{eq:sigma_xy_II}
\sigma_{xy}^{II} = \frac{1}{2}(\lambda^{xy}_+ - \lambda^{xy}_-).
\end{equation}
Using yet one more simple identity 
\begin{equation}
\label{eq:Id1}
\epsilon^{3ij} \mathrm{tr}(\tau_3 \partial_i A_j) 
=\frac{1}{4} \epsilon^{3ij} \mathrm{tr} (Q\partial_i Q \partial_j Q),
\end{equation}
one finally recognizes that the action~(\ref{eq:S_II_top_A}) assumes the explicitly 
gauge-invariant form~(\ref{eq:Q_action_top}) \`{a} la Pruisken.

We finally evaluate the Hall conductivity $\sigma_{xy}^{II}$ in more concrete terms.
Following Burkov \& Balents~\cite{Burkov2011Weyl}, we reinterpret the integration
over $p_3$ in Eq.~(\ref{eq:lambda}) in a manner that cannot be justified from the
linearized two-node approximation alone: let's understand $p_3 + b\equiv m$ as the
masses of a stack of two-dimensional Dirac fermion systems with the quantized
momentum $p_3 = 2\pi n/L_z$. Then
\begin{eqnarray}
\label{eq:sigma_xy_II_sum}
\sigma_{xy}^{II} &=& \frac{2\pi}{L_z} \sum_n \sigma_{xy}^{II,n},
\end{eqnarray}
where $\sigma_{xy}^{II,n}$ is the Hall conductivity of a single 2d layer. Substitution of Eqs.~(\ref{eq:lambda}) and
(\ref{eq:sigma_xy_II}) leads to 
\begin{equation}
    \sigma_{xy}^{II,n}=\frac{im}{\pi }\sum_s s\int_{-\infty}^{\epsilon}  d\epsilon'\int dk\,\frac{1}{[(\epsilon'+i \kappa s)^2-k^2-m^2]^2}.
\end{equation}
(As a side remark, we note the similarity to Eq.~(29) of
Ref.~\onlinecite{Ludwig:1994} where the integer quantum Hall effect was modeled in
terms of two-dimensional Dirac fermions, similar to the fermions populating our
stacked 2d compound insulators.). Doing the momentum integrals, we arrive at
\begin{align}
    \sigma_{xy}^{II, n}&=\frac{im}{4\pi^2 }\sum_s s\int_{-\infty}^{\epsilon} d\epsilon'\frac{1}{(\epsilon'+i \kappa s)^2-m^2}=\frac{C_n}{2\pi}, \nonumber \\
    C_n  &= (1/2\pi)\sum_{\sigma = \pm } \arctan\left[(m+ \sigma \epsilon)/{\kappa}\right].
\end{align}
Here $C_n(b)$ can be interpreted as the disorder averaged Chern number of the $n$-th
layer which depends on $b$ via the ‘effective’ mass $m=2\pi n/L_z +b$. To evaluate
the final sum~(\ref{eq:sigma_xy_II_sum}) which gives us $\sigma_{xy}^{II}$ we note
that only {\it changes} in $C_n$ at zero crossings of the effective mass $m$ can be
unambiguously determined from the linearized theory. Thus we fix the absolute value
of (\ref{eq:sigma_xy_II_sum}) by the condition $\sigma_{xy}^{II} = 0$ at $b=0$. The
latter amounts to replacing $C_n(b) \rightarrow  C_n(b)-C_n(0)$. Then in the limit
$L_z b \gg 1$ we find
\begin{eqnarray}
\label{eq:SigmaXYSingleNode}
\sigma_{xy}^{II} = b/(2\pi),  
\end{eqnarray}
for the contribution of either node $n=1,2$ to the Hall conductivity. (To see how the coupling constants $\sigma_{xy}^{II}$ add in the computation of response coefficients to the full Hall conductivity, expressed in the units of $e^2/h$, see below.) Note that in our linearized model $\sigma_{xy}^{II}$ is independent of both the energy $\epsilon$ and the disorder strength $\kappa$, which is also consistent with the analysis of 
Ref.~\onlinecite{Burkov2014_AHE} in the limit $\epsilon \ll v\Lambda$.

\subsection{Derivation of the Chern-Simons action $S_\mathrm{CS}$}
\label{sec:SCSDer}

The CS action contains terms of order ${\cal O}(A^2 q)$ and ${\cal O}(A^3)$ which produce the first
and the second  contributions to the CS action~(\ref{eq:A_action_CS}), resp. The
first CS term, $S^I_{\rm CS}[A]$, follows from quadratic action $S^{(2)}[A]$. Using Eq.~\eqref{eq:Moyal} to process the ensuing operator products, we obtain
\begin{align}
    &{\cal S}^{(2)}[A] = -\frac{1}{2}\,\mathrm{tr}(\hat G \sla{A} \hat G\sla{A}) \simeq \\
    & -\frac{1}{2}\int dp dx \,\mathrm{tr}\big( (G_{p}\sla{A}-\frac{i}{2}\partial_pG_p \partial_x \sla{A}) (G_{p}\sla{A}-\frac{i}{2}\partial_p G_p\partial_x \sla{A})\big). \nonumber 
\end{align}
Here, the leading contribution here corresponds to Eq.~(\ref{eq:S2_0}). Keeping the
next-to-leading term at order ${\cal O}(A^2 q)$ we obtain
\begin{align}
&{\cal S}^{(2)}_1[A] =
\frac{i}{2}\sum_{ss'}\int dp dx \,\mathrm{tr}\left(P^s \partial_{p} G_p^s \partial_{x } \sla{A} P^{s'}G_{p}^{s'}\sla{A} \right), 
\end{align}
where we  explicitly distinguish between retarded and advanced Green's functions.
At this stage it is advantageous to use the relation $\partial_{p_i} G_p^s = G_p^s \sigma_i G_p^s$. With this trick one arrives at
\begin{align}
&{\cal S}^{(2)}_1[A] = \nonumber
\\& \frac{i}{2}\sum_{ss'}\int dp dx \,\mathrm{tr}\left(P^s G_p^s \sigma_i G^s_p \partial_{x_i} A_j \sigma_j P^{s'}G_{p}^{s'}A_k \sigma_k\right) \nonumber \\
& =  -\sum_{ss'}F_{ss'}\int dx\,\epsilon^{ijk} \,\mathrm{tr}\left(P^s \partial_{x_i} A_j  P^{s'}A_k\right),
\end{align}
where a summation over spatial indices $i,j,k$ is left implicit 
and the integrals $F_{ss'}$ have the form
\begin{equation}
F_{ss'} = \int dp  \,(N_p^s)^2 N_p^{s'} \left( \frac{1}{3} p^2\left(2\epsilon_s + \epsilon_{s'}\right)
- \epsilon_s^2 \epsilon_{s'} \right).   
\end{equation}
Unlike the related Eq.~(\ref{eq:f_ss}) for the coefficients $f_{ss'}$, the above fast momentum integrals are UV finite, and their straightforward evaluation yields
\begin{equation}
F_{ss'} = \frac{1}{8\pi}\left\{
    \begin{array}{ll}
         i s, & \quad s=s' \\
        { 2 \epsilon}/{ ( 3 \kappa )}, & \quad s \neq s'.
    \end{array}
    \right.
\end{equation}
As the intermediate result, the action at order ${\cal O}(A^2 q)$ acquires the form
\begin{eqnarray}
\label{eq:A2_1}
{\cal S}^{(2)}_1[A] &=& {S}^I_{\rm CS}[A] =\nonumber \\ 
    & - & \epsilon^{ijk}\,\frac{\epsilon}{6\pi\kappa}  \int dx \,\mathrm{tr}
    (A_i P^+
    \partial_j A_k P^-).
\end{eqnarray}
The evaluation of the terms of order ${\cal O}(A^3)$ proceeds in a likewise
fashion. They stem from the cubic action ${\cal S}^{(3)}[A]$ in the leading order
Moyal approximation. In Eq.~(\ref{eq:action_reg})) this amounts to neglecting the
slow momenta $q_{1,2}$ in the propagators, i.e. $G_{p+q} \simeq G_p$.
The cubic piece of the action then takes the form
\begin{align}
&{\cal S}^{(3)}_0[A]=\frac{i}{3}\int dq_{1} dq_{2} dp 
\,\mathrm{tr}(G_{p} \sla{A}_{q_1}G_{p} \sla{A}_{q_2}G_p \sla{A}_{-q_1-q_2}) \nonumber \\
& =- \sum_{s_i}\int dx\, F_{s_1 s_2 s_3} \epsilon^{ijk} \mathrm{tr}(P^{s_1} A_{i}P^{s_2}A_j P^{s_3} A_{k}). 
\end{align}
where the ‘triangular’ vertices $F_{s_1 s_2 s_3}$ are expressed via the fast momentum integral
\begin{eqnarray*}
 F_{s_1 s_2 s_3} &=& 
-\frac{i}{3}\int dp\, N^{s_1}_pN^{s_2}_pN^{s_3}_p\, \\
&&\times \mathrm{tr}\Bigl((\epsilon^{s_1}-\sla{p})\sigma_1 
(\epsilon^{s_2}-\sla{p})\sigma_2 (\epsilon^{s_3}-\sla{p})\sigma_3\Bigr)=\\ & = & \frac{2}{3} \int dp \,N^{s_1}_pN^{s_2}_pN^{s_3}_p \\ 
&&\times \Bigl(  \epsilon^{s_1}\epsilon^{s_2}\epsilon^{s_3}-\frac{p^2}{3}(\epsilon^{s_1}+\epsilon^{s_2}+\epsilon^{s_3} )\Bigr).
\end{eqnarray*}
The final evaluation of the (UV finite) momentum integral yields
\begin{equation}
F_{s_1 s_2 s_3}  = \frac{1}{12 \pi}\left\{
\begin{array}{ll}
        +i, &\quad \sum_{i} s_i =+3\\
        -i, &\quad \sum_{i} s_i =-3\\
        {2\epsilon}/{(3\kappa)}, &\quad \sum_{i} s_i =\pm 1.
    \end{array}
    \right.
\end{equation}
Separating now real and imaginary parts, we see that the cubic action ${\cal S}^{(3)}_0[A]$ reads
\begin{equation}
\label{eq:A3_0}
{\cal S}^{(3)}_0[A] = {S}^{II}_{\rm CS}[A] -  
\epsilon^{ijk}\,\frac{\epsilon}{6\pi\kappa}  \int dx \,\mathrm{tr}
    (A_i P^+
    A_j A_k P^-).
\end{equation}
To obtain the final form of the action one has to combine Eqs.~(\ref{eq:A2_1}) and
(\ref{eq:A3_0}). Since  $A$ has the structure of a full gauge, i.e. $A_k =
T^{-1}\partial_k T$, the relation $\epsilon^{ijk} A_j A_k = - \epsilon^{ijk}
\partial_j A_k$ holds, which means that the real parts of the two contributions
cancel each other and we are left with the CS action~(\ref{eq:A_action_CS}).

\emph{Gauge invariance of the CS-action} ---. We conclude this section with a discussion of the gauge
invariance of the CS action. Notice that our initial regularized action $S[Q]$, see
Eq.~(\ref{eq:S_Q_reg}), expressed via the matrix field $Q=T \tau_3 T^{-1}$, is
manifestly invariant under the transformation $T \to T k$, where $k(x)$ is any
spatially dependent matrix from the unbroken part of the symmetry group, $ H$ whose
elements have block-diagonal structure in retarded-advanced space, $k = {\rm bdiag}(
k_+, k_-)^{\rm ar}$, with $k_\pm \in U(R)$. However, in the language of $A$-fields
the invariance under local transformation $k(x)$ is no longer manifest. Recalling the
definition $A_i = T^{-1}\partial_i T$, we find that the $A$'s transform as
non-abelian gauge fields, $A_i \to A_i' = k^{-1} A_i k + k^{-1}
\partial_i k$, and ensuring the invariance of a low-energy field theory under such
gauge transformation is a non-trivial self-consistency check.

The part of the $\sigma$-model's action, lacking explicit gauge invariance, is the CS term~(\ref{eq:CS_AP}). In fact, the gauge transformation specified by the matrix $k(x)$ generates 
a contributions of order ${\cal O}(\partial_i k)^n$ with $n=1,2,3$. Explicit inspection of Eq.~(\ref{eq:SC_diff_form}) shows that
\begin{equation}
S_{\rm CS}[A'] = S_{\rm CS}[A] + \sum_{n=1}^3 \delta S_{\rm CS}^{(n)}[A,k].
\end{equation}
The first two contributions here read as
\begin{eqnarray}
\delta S_{\rm CS}^{(1)}[A,k] &=& -\frac{i}{8\pi}\int \mathrm{tr}\Bigl( \tau_3 \, A \wedge A \wedge k^{-1} dk \Bigr),  \\
\delta S_{\rm CS}^{(2)}[A,k] &=& \frac{i}{8\pi}\int \mathrm{tr}\Bigl( \tau_3\,  A \wedge (k^{-1} dk) \wedge (k^{-1} dk) \Bigr), \nonumber 
\end{eqnarray}
and with the use of a relation $dA = - A\,\wedge\, A$ their sum can be reduced to a boundary action, 
$\delta F=\delta S_{\rm CS}^{(1)} + \delta S_{\rm CS}^{(2)}$, where $\delta F$ is a surface integral over the $2$-form,
\begin{equation}
\delta F[A,k] = \frac{i}{8\pi}  \oint  \mathrm{tr}\Bigl( \tau_3\,  A \wedge k^{-1} dk  \Bigr).
\end{equation}
At first sight, it is not really obvious how to handle this contribution. In other
contexts described by Chern-Simons actions, the fractional QHE, for example, it is
customary to \emph{postulate} a surface action, which is designed so as to cancel the
gauge contributions coming $\delta F$ from the bulk after a gauge transformation.
However, in the present context, both the formal structure, and the physical meaning
of such type of surface contribution remain opaque. Fortunately, the problem has a
much simpler solution. For once (in this section), we need to take into account that
the system has two nodes, and that the other node contains an identical Chern-Simons
action, but of opposite sign. The microscopic analysis of the band structure of Weyl
metals shows that the separation $b$ between the nodes in momentum space vanishes
upon approaching system boundaries; at the boundary, the nodes merge. This means that
the fields $A_n, n=1,2$ describing the nodes effectively hybridize $A_1=A_2\equiv A$
upon approaching the boundary, and this entails the cancellation of the Chern-Simons
actions $S_\mathrm{CS}[A_1]-S_\mathrm{CS}[A_2]\to 0$. With regard to the Chern-Simons sector, our theory therefore remains effectively boundaryless, and the gauge issue does not arise.

The 3rd contribution is solely $k$-dependent and given by the 3d topological $\theta$-term,
\begin{equation}
\delta S_{\rm CS}^{(3)}[k] = \frac{i}{24 \pi} \sum_{s=\pm} s \int \mathrm{tr} ( k_s^{-1}dk_s)^{\wedge 3}.
\end{equation}
This integral gives the quantized value $i\pi (n_+ - n_-)$ where $n_s$ is the winding
number of the configuration $k_s \in U(R)$ in the 3d physical space (recall that
$\pi_3 \left( U(R)\right) = \mathds{Z}$ for $R \geq 2$). We thus see that for large
gauge transformations  the CS action picks up a term quantized contribution $S_{\rm
CS}[A'] = S_{\rm CS}[A]  + i \pi (n_+ - n_-)$. This, however, is not the end of the
story. Recall that the CS action was obtained the gradient expansion of ${\cal
S}_0[A]$ in Eq.~(\ref{eq:action_reg}). It was shown by
Redlich~\cite{Redlich1984Parity} that the regulator action ${\cal S}_\eta[A]$ also
changes under large gauge transformations, and by the same term ${\cal S}_\eta[A'] =
{\cal S}_\eta[A] + i \pi (n_+ - n_-)$. This change results from zero-crossings of
energy levels of the regularized Dirac operator under the spectral flow induced by
$k$. The sum of the two contributions therefore changes by $2\pi i \times
(\mathrm{integer})$, which means that the exponentiated action $\exp(S[A])$ remains
properly gauge-invariant.

Below, it will occasionally be useful to work with this action in the coordinate invariant language of differential forms. Defining the standard CS action as
\begin{equation}
\label{eq:SC_diff_form}
{\cal S}_{CS}[A] = \frac{i}{4\pi} \int  {\rm tr}\Bigl( A \wedge dA + \frac{2}{3} A\wedge A\wedge A \Bigr),
\end{equation}
in terms of the one-form $A = A_i dx^i$, the result~(\ref{eq:A_action_CS}) may be represented as
\begin{equation}
\label{eq:CS_AP}
S_{\mathrm{CS}}[A] = \frac{1}{2}\Bigl( {\cal S}_{CS}[A P^-] - {\cal S}_{CS}[A P^+] \Bigr).
\end{equation}
In passing we noted that the connection of the CS action to the Weiss-Zumino type action describing the three dimensional boundary of the
4d class $\mathrm{A}$ topological insulator was recently emphasized in Ref.~\onlinecite{Zhao_Wang:2015}. This construction underpins the interpretation of a 3d single node Weyl system as the boundary theory of a bulk 4d topological insulator of class $\mathrm{A}$.

\subsection{Renormalization}
\label{sec:Renormalization}

The action as derived above is characterized by three coupling constants: the longitudinal conductivity $\sigma_{xx}$ as a coupling constant of the diffusion term, the Hall conductance $\sigma_{xy}$ multiplying the topological term, and the scattering rate $\tau_n^{-1}$ in front of the inter-node scattering term. (The coupling constant of the CS action is topologically quantized.) Fluctuations will renormalize these coefficients at large distance scales. Speaking of length scales, we need to discriminate between scales larger and smaller than the scale $l_n$ at which the nodes are effectively strongly coupled. Comparing the diffusion term and the inter-node scattering term, we identify this crossover scale as $l_n \sim (\sigma_{xx}\tau_n \nu^{-1})^{1/2}$. For length scales $L<l_n$ the two nodal fields $A_n$ fluctuate independently, and contains two independent contributions $S_n[A_n]$, each containing a diffusion term, a topological term, and a CS term (of opposite sign). For larger scales, the action collapses to $S[A]=S_1[A]+S_2[A]$, where $A\equiv A_1=A_2$ is enforced by the inter-node scattering term, and $S[A]$ contains the diffusion and the topological term (of doubled coupling constant), while the CS actions have canceled out. Let us first concentrate on this latter regime and ask how fluctuations renormalize the two remaining coupling constants $\sigma_{xx},\sigma_{xy}$. This question has been answered in Ref.~\onlinecite{Wang1997Localization} within the context of a field theory study of the layered quantum Hall effect (which, as pointed out above, is described by the same action.) Application of two-loop renormalized perturbation theory led to the flow equations for the dimensionless couplings $g_{\mu \nu }\equiv \sigma_{\mu\nu }L$,
\begin{align}
\label{eq:RGEquations}
\frac{\partial g_{xx}}{\partial \ln L}&= g_{xx}- \frac{1}{3 \pi^4 g_{xx}},\cr
\frac{\partial g_{xy}}{\partial \ln L}&= g_{xy}.
\end{align}
What this tells us is that the dimensionless longitudinal conductance shows Ohmic
behavior $g_{xx}\sim L$, up to weak localization corrections (the second term in the
first equation), which are vanishingly weak in the large distance limit. This is the
scaling behavior of a three dimensional Anderson \emph{metal}: even if the
conductance is initially weak, it grows at large distance scales (while the
conductivity remains unrenormalized.) The second line states that the Hall
conductance, too, shows linearly increasing behavior, this time unaffected even by
weak localization. This implies the conclusion that the Hall conductivity  remains
unrenormalized by quantum fluctuations at a value $ \sigma_{xy}=b/\pi$, twice as
large as the single node contribution~\eqref{eq:SigmaXYSingleNode}. (We here work
under the assumption that the value of the Hall conductivity did not change in the
short distance regime, $L<l_n$, see below.) Notice that the lack of renormalization
of the Hall conductivity distinguishes the system from the genuine 2d quantum Hall
system, in which the Hall conductivity renormalizes towards integer values due to
instanton fluctuations~\cite{Pruisken1984a}.

We finally speculate at what happens at intermediate length scales between the mean
free path $l$ and the crossover scale $l_n$. In this regime, the fields are
effectively decoupled, although the initially weak coupling term is RG relevant (with
engineering dimension $3$), and keeps growing until it becomes of the order of the diffusion
term (engineering dimension $1$) at $L\sim l_n$. The action of the nodal fields $A_n$
is otherwise identical to that discussed above, save for the presence of the CS term.
However, the latter is less relevant than both the diffusion and the topological
term, due to its higher number in derivatives. We therefore suspect that it will not
qualitatively interfere with the renormalization of the theory. This in turn implies
that in the short distance regime, too, the coupling constants renormalize according
to Eqs.~\eqref{eq:RGEquations}. In section~\ref{sec:Keldysh_FT} we will discuss how
the coefficients $\sigma_{\mu \nu}$, here understood as abstract coupling constants, determine the longitudinal and Hall conductance of the system.

Summarizing, we argue that the flow~\eqref{eq:RGEquations} determines the
renormalization of the theory for arbitrary length scales $L>l$. Quantum fluctuations
are generally weak, due to the fact that the dimensionality of the system is larger
than the sigma model lower critical dimension $2$. Physically, this means that the 3d
system behaves as an Anderson metal, whose essentially constant longitudinal
conductivity is determined by the distance of the chemical potential from the nodal
point, which in turn determines the bare $\sigma_{xx}$. Even at the nodal points, the
bare value of $\sigma_{xx}$ obtained from the SCBA/RG analysis of
Sec.~\ref{sec:Saddle_point} is (numerically) larger than than the value
$g_c\, L^{-1}$ with $g_c = 1/\sqrt{3} \pi^2$ marking the 3d Anderson transition towards an insulating phase,
cf. Eq.~\eqref{eq:RGEquations}. While this
observation may not sound too convincing (as mentioned above, the derivation of the
nodal theory is not well controlled parametrically), the fact that a single Weyl node
can be understood as the surface theory of a fictitious 4d topological insulator
implies topological protection against Anderson localization.  However, we
caution that very different things can happen if the system is geometrically confined
to quasi two or one dimensional geometries. This point, too, will be
discussed in Sec.~\ref{sec:Keldysh_FT}.

\section{Gauge field theory}
\label{sec:GFT}

In this section we generalize our  model to the presence of an external  (non-abelian) gauge field $\sla{a} = a_i \sigma_i$. (The non-abelianness is required to describe source fields, see below.) The system then is described by the 
replicated action~(\ref{eq:S}) where the single-particle Hamiltonian incorporates the vector potential,
\begin{equation}
\label{eq:H_a}
 	\hat H[a] = v (\sla{\hat{k}} - \sla{a})\, \sn_3 +(v\sla{b} + b_0)+  V (\bx),
 \end{equation} 
We assume that the fields $a_i$ are structureless in the nodal space --- electrons from different nodes interact with 
the same gauge field --- and (most generally) are elements of the Lie algebra,
$a_i \in \mathfrak{u}(R) \times \mathfrak{u}(R)$, associated with the small symmetry group $H$ (see 
Sec.~\ref{sec:Model} for the definition). 
For practical usage it means that $a_i = \{a_i^{s,rr'}\delta^{ss'}\}$ is a matrix in  replica and
retarded-advanced spaces which commutes with $\tau_3$. In what follows the $a_i$'s will incorporate external sources, and external
physical electric and magnetic fields. Within the Keldysh framework discussed in Sec.~\ref{sec:Keldysh_FT}, the
fields $a_i$ can be also made dynamical to describe particle interactions.
In this section we are primary concerned with the geometrical gauge structure of the theory, which in turn yields far reaching insights into the universal behavior of observables. For the moment, we 
restrict ourselves to the limit $\tau_n \to \infty$ in which the nodes can be discussed separately.

After the integration over fermions,   the  disorder averaged partition function $Z[a]$ becomes a functional of $a$. As in
preceding Sec.~\ref{sec:FTA} we reduce it  
to a path integral over  soft Goldstone fluctuations
\begin{equation}
Z[a]  = \int_{T \in G/H} {\cal D}\, T({\bf x}) \exp(-S[T,a]).
\end{equation} 
The emergent $\sigma$-model action $S[T]=S_d[Q] + S_{\rm top}[Q] + S_{\rm CS}[A]$ 
with $Q=T \tau_3 T^{-1}$ and $A_i = T^{-1}\partial_i T$ describing the theory in the absence of $a$ was constructed above. In the next 
Sec.~\ref{sec:Gauge_inv_action} we show how the general form of $S[T,a]$ can be easily identified on the basis of $S[T]$
using straightforward principles of gauge invariance. 
After that in Sec.~\ref{sec:Saddle_point} we derive the corresponding saddle-point equation of the new theory. 
These equations will be gauge invariant variants of the Usadel equations~\cite{Usadel:1970} introduced long ago in the context of quasiclassical superconductivity. The equations describing the Weyl node are  enriched by contributions stemming from the  
CS terms. In Sec.~\ref{sec:Keldysh_FT} we will see how this leads to essential modifications by which the  diffusive hydrodynamics and the (non-equlibrium) magnetotransport of  disordered Weyl fermions differ from that of conventional fermions. Several of these structures are 'non-perturbative' in that they cannot be obtained from diagrammatic perturbation theory approaches. 

\subsection{Gauge invariant action}

\label{sec:Gauge_inv_action}

Our starting point, the gauge coupled regularized action from which the effective theory is derived, reads  (cf.
Sec.~\ref{sec:Reg}),
\begin{align}
	S[Q,a] &= S_0[Q,a] - S_\eta[Q,a]\\ 
&- \mathrm{tr}\ln\left(i\kappa Q - v\sla{k} 
+ v\sla{a} \right) +\mathrm{tr}\ln\left(i\eta Q - v\sla{k} + v\sla{a} \right). \nonumber
\end{align}
where the self energy $\kappa$, too, may depend on $a$, for example if the latter
describes a very strong magnetic field. However, our focus in the following will be
on regimes where the fields are weak enough for such effects not to play a role. Note
that we have chosen to include the external field $a$ in the regulator $S_\eta$.
After the similarity transformation the  action takes the form
\begin{eqnarray}
\label{eq:det_a_reg}
	S[Q,a]&=&  {\cal S}[\bar A] - {\cal S}_\eta[\bar A]; \\
{\cal S}[\bar A]&=&-\mathrm{tr}\ln\left(i\kappa \tau_3  - v\sla{k}+ i v \sla{A} + v T^{-1} \sla{a} T \right), \nonumber \\
{\cal S}_\eta[\bar A] &=& - \mathrm{tr}\ln\left(i\eta \tau_3  - v\sla{k} + i v \sla{A} + v T^{-1} \sla{a} T\right). \nonumber
\end{eqnarray}
We observe that the only difference to Eq.~(\ref{eq:action_reg}) is that 
the diffusion field $A_k$ has been replaced by the (left) gauge invariant combination
\begin{equation}
\label{eq:A_bar}
 \bar A_k \equiv A_k - i\, T^{-1} a_k T = T^{-1} (\partial_k - i a_k )T.
\end{equation}
This expression suggests an interpretation of $\bar A_k$ as the field $a_k$ gauge transformed by $T$. Indeed, under a transformation   $\psi \to \psi' = T \psi$ of the fields entering the native fermion action 
the naked  $a_k$ changes to the gauge transformed configuration~\eqref{eq:A_bar}. The same argument tells us that in our previous discussions we have been working with fields $A_k = T^{-1}\partial_k T$ that were 'pure gauges'. It also suggests to take a look at the field strength tensor 
\begin{equation}
\label{eq:F_jk}
\bar F_{jk} = \partial_j \bar A_k - \partial_k \bar A_j
+ [\bar A_j, \bar A_k].
\end{equation}
For $a=0$, the field strength vanishes, as one expects for a pure gauge. However, for the configurations~\eqref{eq:A_bar} one finds $\bar F_{jk} = -i T^{-1} f_{jk} T$, where
\begin{equation}
\label{eq:f_jk}
f_{jk} = \partial_j a_k - \partial_k a_j
- i [a_j, a_k]
\end{equation}
is the field-strength tensor of $a_i$. In the following we assume that $f_{jk}$ is proportional to 
the identity matrix in the retarded-advanced and replica space, i.e. 
$f_{jk} \propto \mathds{1}_{ra} \otimes \mathds{1}_R$. 
This  assumption leaves us enough freedom to treat all 
important cases. Before proceeding, we remark that a consistent regularization of the theory requires a coupling of the regulator action to $a$. This can lead to side effects, viz. a dependence of the \emph{regulates} action on non-analytic functions of the field strength tensor~\cite{Redlich1984Parity}. This --- conceptually interesting --- point is discussed in Appendix~\ref{sec:RegulatorField} where we show that those terms drop out under the conditions formulated above.

The  $\sigma$-model describing a single copy of disordered Weyl fermions in the presence of the field $a$ is described by the action
\begin{equation}
\label{eq:action_Ta}
S[T,a] = S_{\rm d}[Q,a] +  S_{\rm top}[Q,a] + S_{\rm CS}[\bar A].
\end{equation} 
Here the diffusive and AHE terms are the canonical gauge-invariant extensions of 
the actions~(\ref{eq:Q_action_diff}) and (\ref{eq:Q_action_top}):
\begin{eqnarray}
\label{eq:Q_diff}
S_\mathrm{d} [Q,a] &=&
\frac{\sigma_{xx}}{8}\int dx\,\mathrm{tr}(\nabla Q)^2, \\
\label{eq:Q_top}
S_\mathrm{top}[Q,a] &=&\frac{\sigma_{xy}}{8}
 \epsilon^{3ij}\int dx\,\mathrm{tr}
(Q\nabla_i Q \nabla_j Q),
\end{eqnarray}
where the $\nabla$ operator denotes the long derivative,
\begin{equation}
\nabla_k Q := \partial_k Q - i [a_k,Q].
\end{equation}
The CS contribution to the action has the same functional form~(\ref{eq:A_action_CS}) as discussed before
but is evaluated with the use of the gauge-invariant superposition $\bar A_k$ of the fields $A_k$ and $a_k$ defined
by Eq.~(\ref{eq:A_bar}). 

Under a left gauge transformation $U$ acting on the fields as $T' = U T$ fields and derivatives change as
\begin{equation}
 Q' = U Q U^{-1}, \quad \nabla'_i Q' = U (\nabla_i Q) U^{-1} 
\end{equation}
(where $\nabla_k' = \partial_k - i[a_k', \cdot ]$), and this shows that
each piece of the action~(\ref{eq:action_Ta}) is  (left) gauge invariant.

We finally note  another equivalent representation of the CS action,
\begin{eqnarray}
\label{eq:SC_tau}
S_\mathrm{CS}[\bar A] = &-& \frac{i}{16\pi}\int dx \,\epsilon^{ijk} \mathrm{tr} \Bigl(2 \partial_i \bar A_j \tau_3 \bar  A_k   \nonumber \\
& + &\bar A_{i}   \bar A_{j}  \tau_3  \bar A_{k} 
  + \frac{1}{3} \tau_3 \bar A_{i}   \tau_3 \bar A_{j}  \tau_3 \bar A_{k}\Bigr).
\end{eqnarray}
which will be useful throughout.

As mentioned above, the structure of the above action follows essentially from principles of gauge invariance. However, for the sake of completeness, some more details on its construction are included in Appendix~\ref{sec:Gauge_inv_action}.

\subsection{Kinetic equations}

\label{sec:Saddle_point}
 
Let us now start from the  gauged form of the $\sigma$-model to derive the
saddle-point equations of the theory. Their form is interesting in its own right and
will be used in Sec.~\ref{sec:Keldysh_FT} for the analysis of the magnetotransport in
the system.

To identify stationary configuration $Q = T\sigma_3 T^{-1}$ of the action~(\ref{eq:action_Ta})  we may parametrize  fluctuations around $Q$ as
$T' = e^\lambda T$ with $\lambda \ll 1$. This defines the variations 
\begin{eqnarray}
\label{eq:variations}
\delta Q &=& [\lambda, Q], \qquad   \\
\delta A_i &=&   - (T^{-1} \delta T) A_i + T^{-1} \partial_i\delta T, \nonumber
\end{eqnarray}
where $\delta T = \lambda\, T$. Substituting the fields $Q' = Q + \delta Q$ and $A'_i = A_i + \delta A_i$ into
the action and requiring terms linear in $\lambda$  to vanish one obtains the saddle point equation 
\begin{align}
\label{eq:Usadel_a}
&  D \nabla_{k} (Q \nabla_k Q) 
- \frac{ \epsilon^{ijk}}{32\pi^2\nu} \Bigl[ \{ f_{ij}, Q \}, Q \nabla_k Q \Bigr]  \nonumber \\
& \qquad - \frac{i \epsilon^{ijk} }{16\pi^2\nu} 
(\nabla_i Q) ( \nabla_j Q )(\nabla_k Q) = 0.
\end{align} 
We have introduced here the diffusion coefficient $D$ 
and the DoS at the Fermi surface $\nu$ which are related to the conductivity $\sigma_{xx}$ per node 
by the familiar Einstein relation $\sigma_{xx} = 2 \pi \nu D$.
For  details on the  derivation of this central result 
we refer to Appendix~\ref{app:Usadel}.

Equations of this type are widely known in the context of mesoscopic
superconductivity, where they are known as Usadel equations~\cite{Usadel:1970}. More
generally, these equations belong to the family of kinetic equations, and we will
refer to them as such throughout. Our kinetic equation~(\ref{eq:Usadel_a}) is
explicitly gauge invariant --- it holds for the $Q$-field as such, rather than for
the gauge fields $A_k$. Its first term stems from the action $S_d$~(\ref{eq:Q_diff}),
while the two others originate in the CS action (the topological AHE
action~(\ref{eq:S_top_contour}), which can be reduced to a pure boundary term, does
not contribute to the saddle point equation in the bulk). Finally, the interpretation
of the  drift term containing only one derivative is most naturally revealed in
quasi-one-dimensional geometries, as discussed in details in Sec.~\ref{sec:MC_noise}.
We here merely state that this term describes the CME in its most general
manifestations.

To establish the meaning of the last third term in Eq.~(\ref{eq:Usadel_a}), we temporarily set $a_i=0$, write the term in an invariant form, $\propto \frac{1}{16\pi} \mathrm{tr}(dQ\wedge dQ\wedge dQ)$,  and integrate 
over a   volume $\Omega$. Upon application of Stoke's theorem, the term then assumes the form of a surface integral over the two-dimensional bondary of $\Omega$,
\begin{equation}
W = \frac{1}{16\pi} \int_{\partial \Omega} \mathrm{tr} ( Q\, d Q \wedge d Q ).  
\end{equation}
This integral is topologically quantized and describes the windings $W$ of the field over the boundary. More precisely speaking,  
the 2nd homotopy group of our  $\sigma$-model manifold is non-trivial, $\pi_2 (G/H) = \mathds{Z}$, and the integral measures the topological contents of fields. However, the presence of such surface windings, necessitates the presence of singular topological defects in the bulk of $\Omega$, the simplest example being a `hedgehog' configuration. Whether or not the inclusion of such structure in the field theory of the
disordered Weyl system is of any importance requires  further study.
(There is the intriguing observation that singular vortex-like topological excitations  of \emph{two}-dimensional $\sigma$-models play a crucial
role in the description of Anderson localization transitions in chiral~\cite{Koenig:2012} and symplectic~\cite{Kane-Fu2012}
symmetry classes.)

Importantly, the generalized kinetic equation  can be cast in the form of a continuity equation.
Define the matrix current $J^i = J^i_{\rm D} +  J^i_{\rm CS}$ to be a sum of diffusive ($J^i_{\rm D}$) 
and CS ($J^i_{\rm CS}$) currents, resp.
\begin{eqnarray}
\label{eq:D_current}
J^i_{\rm D} &=&  \pi \nu D (Q \nabla_i Q) , \\
\label{eq:CS_current}
J^i_{\rm CS} &=& - \frac{i\epsilon^{ijk} }{16\pi }\Bigl( Q ( \nabla_j Q )(\nabla_k Q) - i \{f_{jk}, Q\}\Bigr). 
\end{eqnarray}
The saddle point Eq.~(\ref{eq:Usadel_a}) then assumes the form
\begin{equation}
\label{eq:dQ_divJ}
\nabla_i J^i = 0. 
\end{equation}
This structure follows from general properties of covariant 
derivatives (the $\nabla_i$'s): for arbitrary matrices $X, Y$ one has the `chain' rule
\begin{equation}
\nabla_k ( X Y) = (\nabla_k X) Y + X \nabla_k Y, 
\end{equation}
while two derivatives along different directions generally do not commute,
\begin{equation}
\left[\nabla_j, \nabla_k\right] X
= - i [f_{jk}, X], 
\end{equation}
if the connection $a_i$ is not the full gauge. We further have the Bianchi identity, $\nabla f=0$, which in coordinates reads as 
\begin{equation}
\epsilon^{ijk} \nabla_i f_{jk} = 0,
\end{equation}
With this information in store, we find
\begin{align}
\nabla_i J^i_{\rm CS} & = - \frac{i\epsilon^{ijk} }{16\pi}
\Bigl( \nabla_ i Q \nabla_j Q \nabla_k Q  \nonumber \\
&- \frac i2 Q [ f_{ij}, Q] \nabla_k Q - \frac i2 Q  \nabla_j Q  [f_{ik}, Q] \nonumber \\
&- i \{f_{jk}, \nabla_i Q\}\Bigr).
\end{align}
Using the   identity $Q \nabla_k Q  = - ( \nabla_k Q ) Q$ 
one can  regroup the terms above to obtain the divergence of the CS current,
\begin{equation}
\nabla_i J^i_{\rm CS} = -\frac{i\epsilon^{ijk} }{16\pi} \left( \nabla_i Q \nabla_j Q \nabla_k Q 
- \frac i2 \Bigl[ \{ f_{ij}, Q \}, Q \nabla_k Q \Bigr] \right),
\end{equation}
and hence the equivalence of the kinetic~(\ref{eq:Usadel_a}) and the
continuity~(\ref{eq:dQ_divJ}) equations is established.

In accord with  general principles -- gauge invariance implies current conservation
-- the continuity equation~(\ref{eq:dQ_divJ}) discussed above emerges from the local
non-Abelian (left) gauge symmetry of the $\sigma$-model action~(\ref{eq:Usadel_a}).
Indeed, by construction (see Sec.~\ref{sec:Gauge_inv_action}) the action $S[T,a]$ is
invariant under the simultaneous variations of the fields $T$ and $a$ in the form
\begin{equation}
\delta T= \lambda\, T, \qquad \delta a_k = [\lambda, a_k] - i \partial_k \lambda,
\end{equation} 
resulting from the infinitesimal rotation $U = e^\lambda \simeq 1 + \lambda$. In more explicit terms we have
\begin{align}
\delta S &=\int d x \, {\rm tr} \left\{  \lambda \,\left(\frac{\delta S}{\delta \lambda}\right)_a + 
\delta a_k \, \frac{\delta S}{\delta a_k}\right\} 
\nonumber \\
&= 
\int d x\, {\rm tr}\, \lambda \, \left\{\left(\frac{\delta S}{\delta \lambda}\right)_a 
 + i \nabla_k \left(\frac{\delta S}{\delta a_k}\right) \right\}  = 0,
\end{align}
where $\left({\delta S}/{\delta \lambda}\right)_a$ denotes the variation of $S$ under fixed gauge field $a$.
Since in the last equation $\lambda$ is arbitrary, we find the following identity
\begin{align}
&\left( \frac{\delta S}{\delta \lambda}\right)_a  +  \nabla_k J^k = 0, \\
\label{eq:J_from_S}
&\qquad J^k = i\frac{\delta S}{\delta a_k}.
\end{align}
which is valid for any $Q$-matrix and vector potentials $a$. On the saddle point
level the $Q$-matrix is the solution to $\left( {\delta S}/{\delta \lambda}\right)_a
= 0$, which is the implicit form of the kinetic equation~(\ref{eq:Usadel_a}). Hence we
see that the latter should be equivalent to the continuity relation $\nabla_k J^k =
0$. Invoking the explicit form of the action $S[T,a]$ one then checks that the
current~(\ref{eq:J_from_S}) is but the sum of diffusive~(\ref{eq:D_current}) and
Chern-Simons~(\ref{eq:CS_current}) currents postulated above.

\section{Keldysh field theory}

\label{sec:Keldysh_FT}

Our so far discussion was mainly concerned with internal structures of the theory, which are best exposed within the simple replica framework, at fixed values of Green function energies. In
this section, we will go beyond this level to explore the physics of observables. Our discussion will include
non-equilibrium observables, and this requires extension of the theory to a dynamical (time dependent) framework, which in the present context is the real-time Keldysh field theory approach. We start in Sec.~\ref{eq:real_time} by discussing the straightforward reformulation of both the 
$\sigma$-model action and the derived 
kinetic equation within the Keldysh framework. After that, in Sec.~\ref{sec:Transport} we will proceed to the  discussion of various features of the Weyl
semimetal, highlighting  manifestations of the axial anomaly, and structures which are not straightforwardly obtained by other methods. Specifically, in
Sec.~\ref{sec:hydrodynamics} we derive the diffusive hydrodynamics of the two node
system, and Sec.~\ref{sec:MC_noise} is devoted to the study of the magnetoconductance
and of non-equilibrium shot noise.

\subsection{Real-time nonlinear $\sigma$-model}
\label{eq:real_time}

\subsubsection{Effective action and kinetic equations}
\label{sec:Usadel_K}

Given the replicated $\sigma$-model action at fixed energy $\epsilon$, which was
derived and discussed in details in Sec.~\ref{sec:FTA}, the extension to the
Keldysh-like theory framework follows the lines of the general framework formulated within the context of interacting diffusive metals~\cite{Kamenev:1999} and
superconductors~\cite{Feigelman:2000}, see also the excellent
text~\cite{Kamenev:book} for a pedagogical introduction to the field.

One starts from the real-time analog of the Gaussian action~(\ref{eq:S})
\begin{equation}
\label{eq:S_real_rime}
	S[\bar \psi,\psi]=-i\int_{{\cal C}_K} dt\, dx \,\bar \psi(i\partial_t - \mu - \hat H)\psi,
\end{equation}
where the time integral is performed along the Keldysh contour (${\cal C}_K$) and
$\mu$ is the chemical potential. The field, which now depends on real time and space,
$\psi = (\psi^f_{i,n}(t,{\bf x}), \psi^b_{i,n}(t,{\bf x}))^T$ becomes a spinor defined
by the 4-component fields $\psi^{f(b)}$ residing on the forward (backward) branch of the path ${\cal C}_K$, where the four components
represent the nodal ($n=1,2$) and the Weyl spinor ($i=1,2$) structure
of the model Hamiltonian $\hat H$ as before.  When Fourier transformed, the fields $\psi^{f/b}_{\epsilon,i,n}({\bf x})$ 
acquire an energy index $\epsilon$ (replacing time) which under discretization plays a role analogous 
to the index $r=1,...,R$ of the old replica theory, while the advanced/retarded index $s$ is replaced by
the forward/backward structure of the original Keldysh spinor after a linear transformation, 
see details in Refs.~\cite{Kamenev:review, Kamenev:book}.

Unlike with the  replica theory, the physical stationary saddle point 
$\bar B= - i \kappa \hat \Lambda$ 
of the effective Keldysh action is not diagonal in retarded-advanced subspace. It 
depends on the electron distribution function
$h_\epsilon \equiv 1 - 2f_\epsilon$ as
\begin{equation}
\Lambda_{\epsilon \epsilon'} = \left( 
\begin{array}{cc}
1 & 2 h_\epsilon \\
0 & -1
\end{array}
\right)_{\rm RA}\!\delta_{\epsilon \epsilon'}.
\end{equation}
In  equilibrium $f_\epsilon$ is the Fermi distribution leading to $h^F_\epsilon \equiv \tanh (\epsilon/2T)$,
which in the time domain is given by $h^F(t_1-t_2) = -i T/\sinh (\pi T (t_1-t_2))$.
Under more general out-of-equilibrium situation $h=h(t_1,t_2)$ becomes a two-time function which then translates into a non-diagonal matrix $h_{\epsilon \epsilon'}$ in the energy domain.

To establish the connection to the replica field theory, one diagonalizes the physical saddle point $\hat \Lambda$  as follows
\begin{equation}
\label{eq:QDist}
\hat \Lambda = T_0 \circ \tau^{\rm RA}_3 \circ T_0^{-1}, \quad  T_0 = \left( 
\begin{array}{cc}
1 &  h_\epsilon \\
0 & -1
\end{array}\right).
\end{equation} 
Where we use the  $\circ$-symbol to indicate that the diagonalization implies multiplication in energy or convolution in time, depending on the chosen representation.   The soft modes in the Keldysh non-linear $\sigma$-model are then  parametrized as 
\begin{equation}
Q = T \circ \tau^{\rm RA}_3 \circ T^{-1}, \quad T = T_0 \circ T', 
\end{equation}
where the matrices $T'$ --- similar to the replica theory --- span the coset space $ U(2N)/U(N)\times U(N)$ where
$N \to \infty$ is the number of discretization time points in the Keldysh path integral. 
 
The construction of the effective action above was based solely on the coset structure, but not on the detailed realization of the matrices $T$. This means that both, the
$\sigma$-model action~(\ref{eq:action_Ta}) and the kinetic
equation~(\ref{eq:Usadel_a}) retain their form within the Keldysh field framework. At
this stage we choose the gauge field, ${\bf a}=\{{\bf a}^+, {\bf
a}^-\}$, to be the physical electromagnetic field with ${\bf E} = - \partial_t {\bf
a}$ and ${\bf B} = \nabla \times {\bf a}$, where ${\bf a}^\pm$ stand for
the vector potentials on the two branches of the Keldysh contour. (Except for in final formulas, $e=c=\hbar=1$ throughout.) The Keldysh action for
the two-node Weyl metal  then has the form
\begin{align}
\label{eq:S_Keldysh}
&S[T, a] = \nonumber\\
& \sum_{n=1,2} \Bigl( S_{\rm d}[Q_n,a] + S_{\rm top}[Q_n,a] +  (-1)^{n+1} S_{\rm CS}[T_n,a] \Bigr) 
\nonumber\\
&\quad + S_{\rm imp}[Q_1,Q_2] - \frac{i}{8\pi} \int dt \, d x 
\left[(\dot{{\bf a}}^+)^{2} - (\dot{{\bf a}}^-)^{2}\right], 
\end{align}
where the last term is the electrostatic energy with a density ${\bf E}^2/8\pi$. (The magnetic field does not contribute significantly to the overall field energy.) In~(\ref{eq:S_Keldysh}), $S_{\rm top}$, $S_{\rm CS}$, and $S_{\rm imp}$ read as before, see Eqs.~(\ref{eq:Q_top}) 
and (\ref{eq:SC_tau}), while the diffusive part
\begin{equation}
S_{\rm d}[Q] = \frac{\pi \nu}{4}\int dtdx\,{\rm tr}\left(  D ({\nabla} Q)^2 - 4\partial_t Q \right)
\end{equation}
includes an additional dynamical contribution~\footnote{ The latter follows from the
gradient expansion of the regularized action since in the Keldysh field theory the
two-time matrix field $T_{t_1, t_2}(x)$ and the energy operator $\hat \epsilon \equiv
-i\partial_t $ do not commute.}. Finally, the inter-node scattering action is given by the straightforward extension of Eq.~\eqref{eq:Simp}, i.e.
\begin{align*}
S_{\rm imp}[Q_1, Q_2] = - \frac{\pi \nu}{4\tau_n}\int dt d x \, {\rm tr} (Q_1 Q_2)
\end{align*}

The kinetic equation corresponding to the effective acion~(\ref{eq:S_Keldysh}) is obtained by variation of the fields $Q$, as in the replica case. For our choice of the external vector potential, we find 
\begin{equation}
\label{eq:Usadel_K}
\nabla_k J_n^k - \pi \nu [\partial_t, Q_n] +(-)^n  \frac{\pi\nu}{4\tau_n}[Q_1, Q_2] = 0, 
\end{equation}
where the matrices are written in the time domain, $Q_{t_1 t_2}(x)$, 
and the divergence of the matrix current in the node $n$ assumes the form
\begin{eqnarray}
\label{eq:div_Current_K}
\nabla_k J_{n}^k &=& \pi\nu D \nabla_k( Q_n \nabla_k Q_n) +  
\frac{(-1)^n}{4\pi } B_k \nabla_k Q_{n} \nonumber \\
&+& (-1)^n
\frac{i \epsilon^{ijk} }{16\pi} 
(\nabla_i Q_{n}) ( \nabla_j Q_{n} )(\nabla_k Q_{n}).
\end{eqnarray}
We will make extensive use of this equation in our analysis of the transport physics below.

\subsubsection{Drift diffusion dynamics}
\label{sec:DriftDiffusion}

Before turning to the quantitative evaluation of the Keldysh theory, we would like to
 draw the attention of the reader to the  term  $\propto ({\bf B} \cdot {\bf \nabla})
 Q_n$ in Eq.~\eqref{eq:div_Current_K}, which is a direct descendant of the CS term.
 The most interesting feature of this term is that it contains only one derivative.
 Alluding to the similarity of the kinetic equation with a Fokker-Planck equation, we
 anticipate that this is a \emph{drift term}, which describes directed motion of
 charge carriers in a direction specified by the magnetic field. The fact that this
 term contains less derivatives than the diffusion term also means that it will
 dominate the dynamics at large scales. I.e. we are led to the anticipation that
 diffusion at short scales gives way to effectively ballistic motion at large scales.
 This phenomenon is a direct consequence of the CME and not realized in other forms
 of disordered electronic matter. We will discuss various of its concrete
 manifestations below.

To reveal the physical origin of the  drift term we consider the spectrum $\epsilon_z(k_z)$ 
of a clean Weyl node in the presence of a magnetic field ${\bf B}$ of strength $B$ in $z$-direction.
It is well known and is given by
\begin{equation}
\epsilon_n(k_z) = v\, {\rm sgn}(n)\sqrt{ 2|n| B + k_z^2}
\end{equation}
for $n \neq 0$ Landau levels (LL) while the `chiral' $n=0$ LL has the  linear dispersion 
relation 
\begin{equation}
\epsilon_0(k_z) = (-1)^n v k_z,
\end{equation}
where the sign $(-1)^n$ reflects the chirality of each Weyl nodal point. These LLs
are multiply degenerate, where the degeneracy factor $N_\phi $ is given by the number
of flux quanta piercing the cross section $\mathcal{A}$ of the systems transverse to
the magnetic field, i.e. $N_\phi=\mathcal{A}/(2\pi l_B^{2})$, where $l_B= 1/\sqrt{B}$ is
the magnetic length. Each of these states propagates in $z$-direction with a
characteristic velocity $\partial_{k_z}\epsilon_0(k_z)=(-1)^n v$, i.e. there is a
collective drift current density of magnitude $(-)^n v N_\phi \mathcal{A}^{-1}\sim
(-)^n v  B$.

Our kinetic  equation~(\ref{eq:Usadel_K}) states that these multiply degenerate one-dimensional channels 
survive  the presence of  {\it itra}-node impurity scattering. 
Within the theory of quasiclassical
superconductivity~\cite{Eilenberger:1968, Rammer:1986, Muzykantskii:1995, Shelankov:2000}, the propagation of quasi-particles with velocity $v$ along ballistic trajectories locally described by a unit vector $\mathbf{n}$ is described by terms of the structure $ \propto v ({\bf n} \cdot {\bf \nabla}) Q$ in the evolution equations. Normally, such terms arise a length scales \emph{shorter} than the crossover scale to a diffusion regime. Our result above implies the stability of the ballistic drift term at large length scales, the presence of impurity scattering notwithstanding. 

To make the above arguments a little more concrete, consider the
quasi-one-dimensional (1D) geometry (Fig.~\ref{fig:1DCMESchematic}) with a mesoscopic
wire made of Weyl semimetal which has a crossection ${\cal A}$ satisfying the
condition $D/{\cal A} \lesssim {\rm max}\{\epsilon, T\}$. In this case the last term
with triple gradients in Eq.~(\ref{eq:div_Current_K}) can be neglected. Upon
multiplying the kinetic Eq.~(\ref{eq:Usadel_K}) by the doubled area $2 \times {\cal
A}$ we obtain, say, for the 1st node
\begin{align}
\label{eq:KineticQuasi1}
& g_{zz} \nabla_z( Q_1 \nabla_z Q_1) - N_\phi \, \nabla_z Q_1  + 
 \frac{\pi \nu_1}{ 2\tau_n}[Q_1, Q_2] \nonumber \\
&  = 2\pi \nu_1 [\partial_t, Q_1],
\end{align}
where $g_{zz} = 2\pi \nu {\cal A} D$ is the conductance per length of the wire,
$\nu_1 = {\cal A} \nu$ is the one-dimensional DoS, and
$N_\phi = {\cal A} /2\pi l_B^2 = \Phi/\Phi_0$ is the number of magnetic flux quanta
through the wire's cross section. We now recognize that this last equation contains
both a  diffusive term describing the $n\neq 0$ LLs mixed by disorder a drift term
due to $N_\phi$ ballistic 1D channels originating from the $n=0$ LL. Finally, notice
that the dominance of the drift term at large length scales suggest the existence of
a regime of effectively ballistic transport, in which the conductance is proportional
to the number of conducting channels, $N_\phi \propto B$ and independent of the
length of the wire. While the existence of this regime follows from straightforward dimensional analysis, we will see in the next section, that it is easily missed in linear response approaches to transport. 

To conclude this section we mention that the Keldysh action and kinetic equation
discussed above are valid in the limit of sufficiently weak classical magnetic fields corresponding to
$l_B \gtrsim (k_F^{-1} l)^{1/2}$ with $k_F \equiv \mu/v$ when the oscillations in the density of states 
due to Landau quantization smeared by disorder 
are not important (see Ref.~\onlinecite{Klier:2015} for the detailed study of the 
transversal magnetoresistance in the opposite limit of strong magnetic fields).

\subsubsection{Quadratic action}
\label{sec:LinearResponse}

In this section, we will establish contact to previous linear response approaches to transport in the dirty Weyl system~\cite{Burkov2014Chiral,Son:2013}. This can be done without reference to the full  framework of the kinetic equations. What we need to do, rather, is expand the action to quadratic order in the generators $d$ of diffusion modes. The latter are defined by parameterizing the $T$-matrices of node $n$ 
as $T_n = T_0 \circ e^{-W_n/2}$, 
where $\{W_n,\tau_3\} = 0$ is off-diagonal,
\begin{equation}
W_{n,tt'} ({\bf r}) = \left(
\begin{array}{cc}
0 & d_{n,tt'}({\bf r}) \\
\bar d_{n,tt'}({\bf r}) & 0 \\
\end{array}\right),
\end{equation}
in Keldysh advanced/retarded space, and the overbar denotes complex conjugation.
Physically, the fields $d_{n,tt'}(\mathbf{r})$ describe the phase coherent
propagation of an advanced and a retarded quasiparticle amplitude of energy/momentum
close to node $n$. The condition of phase coherence requires to propagating
amplitudes to stay close in space, $\mathbf{r}$, while the times, $t,t'$, at which
$\mathbf{r}$ is traversed can be different.  By definition, the quadratic action
governing the fluctuation of the fields $d$ in a Gaussian approximation to the theory
defines the effective diffusion mode of the system.

Defining a nodal doublet, $d\equiv (d_{n=1},d_{n=2})^T$, the diffusive ($S_d$) and impurity ($S_{\rm imp}$) contribute to the action as
\begin{align}
&S_{\rm d} \overset{{\cal O}(d^2)}{\longrightarrow } - \frac{\pi\nu}{2}\sum_{\epsilon,\epsilon',{\bf q}}
\bar d_{\epsilon\epsilon'}^T({\bf -q}) \left[ D{\bf q}^2 + \omega \right]
d_{\epsilon'\epsilon}({\bf q}), \nonumber \\
&S_{\rm imp} \overset{{\cal O}(d^2)}{\longrightarrow } -\frac{\pi\nu}{4 \tau_n}  \sum_{\epsilon,\epsilon',{\bf q}} 
\bar d_{\epsilon\epsilon'}^T({\bf -q}) [1-\sigma_1] d_{\epsilon'\epsilon}({\bf q}).
\end{align}
where $\omega = \epsilon -\epsilon'$, we have switched to the energy-momentum domain, and Pauli matrices $\sigma_i$ act in node space throughout.

Turning to the CS-term $S_{\rm CS}[\bar A]$, we have  
 $\bar A_i = A_i - i a_i$ where the external vector potential  
$a = B( -y, x, 0)/2$ produces the magnetic field along $z$-axis. 
To quadratic order in $d$ only the first part of the CS action contributes and we find 
\begin{align}
S_{\rm CS}^I[A]& \overset{{\cal O}(d^2)}{\longrightarrow }
-\frac{ B}{4\pi}  {\rm tr} (A_3 \tau_3)\longrightarrow\nonumber\cr
&\longrightarrow - \frac{\pi\nu}{2}\sum_{\epsilon,\epsilon',{\bf q}}
\left( \frac{ i  B} { 4 \pi^2 \nu} \right)\bar d^T_{\epsilon\epsilon'}({\bf -q})  (\sigma_3 q_z) 
d_{\epsilon'\epsilon}({\bf q}).
\end{align}
where in the second line we have expanded the field $A_3 = T^{-1} \partial_3 T\to \frac{1}{8}[\partial W,W]$ to second order in the  generators $W$.

At this stage it is convenient to perform an orthogonal rotation  
in nodal space to symmetric ($d_s$) and axial $(d_a)$ diffusion fields,
\begin{equation}
d_{s/a} = \frac{1}{\sqrt 2}(d_1 \pm d_2).
\end{equation}
Collecting terms, we find that the Gaussian action governing the  doublet $d=(d_{\rm s}, d_{\rm a})^T$ takes the form
\begin{equation}
S =  -\frac{\pi\nu}{2}\sum_{\epsilon,\epsilon',{\bf q}}
\bar d_{\epsilon\epsilon'}^T({\bf -q}) {\cal D}^{-1}(\omega,{\bf q}) d_{\epsilon'\epsilon}({\bf q}),
\end{equation}  
Here ${\cal D}^{-1}(\omega,{\bf q})$ is the (inverse) {\it diffuson},
\begin{equation}
\label{eq:D}
{\cal D}^{-1}(\omega,{\bf q}) = 
\left[ 
\begin{array}{cc}
D{\bf q}^2 - i\omega  &     - i \Gamma q_z \\
  -  i \Gamma q_z &  D{\bf q}^2 - i\omega  + {1}/{\tau_n}
\end{array}
\right],
\end{equation}
and 
\begin{equation}
\Gamma = \frac{  B} { 4 \pi^2 \nu }.
\end{equation}
will be referred as the `drift velocity' in what follows.
The diffusion mode~(\ref{eq:D}) equals that found by diagrammatic perturbation theory in Ref.~\cite{Burkov2014Chiral}.

The physical meaning of $D$ becomes evident once we introduce the total 
($\rho = \rho_+ +  \rho_-$) and the axial ($\rho_{\rm a} = \rho_+ - \rho_-$) electron densities. Eq.~(\ref{eq:D}) for the propagator then implies~\footnote{Formally, this statement is verified by generating the exectation values of $\rho, \rho_a$ by differentiation w.r.t. a source field, and evaluating the sourced functional integral in a Gaussian approximation. However, we do not discuss the straightforward technicalities of this operation here.}  that the two component vector of densities $(\rho,\rho_a)$ evolves according to the equation $D^{-1}_{ij}\rho_j=0$, or
\begin{eqnarray}
\label{eq:n}
 &&- D \nabla^2 \rho + \partial_t \rho - \Gamma \partial_z \rho_{\rm a} = 0, \nonumber \\
 && - \Gamma \partial_z \rho - D \nabla^2 \rho_{\rm a} + \partial_t \rho_{\rm a} + \rho_{\rm a}/\tau_n = 0. 
\end{eqnarray}
Based on these equations and the Einstein relation, Refs.~\cite{Burkov2014Chiral,Son:2013} argued that the longitudinal conductance of the system reads
\begin{equation}
\label{eq:sigma_B}
\sigma_{zz}(B) = \sigma_{xx} + \frac{\tau_n}{2\nu}\left(\frac{ B}{2\pi^2 }\right)^2,
\end{equation}
where the $B$-dependent term is a manifestation of the CME in diffusive transport.
While it is suppressed by inter-node scattering, it persists for any finite $\tau_n$,
which, in fact, poses a problem: one may wonder what cuts off the formal divergence of the
result in the case of uncoupled nodes $\tau_n\to \infty$ and/or strong magnetic field
$B\to \infty$. Our discussion of section~\ref{sec:DriftDiffusion} in fact suggests
that the correct asymptotics should be a result $\sigma_{zz}\propto B$ linear in $B$
and finite for $\tau_n\to \infty$.

In the following we will argue that the derivation of a generalized result which
comprises all limiting cases must include the external electric field driving the
current (which was tacitly ignored in the discussion above), and the back-action of
the latter on the nodal distribution functions $f_{n}$. This information is
conveniently encoded in the kinetic equations, which not only know about the soft
excitations of the system (above parameterized via the $d$-fields), but also about
the feedback of diffusion into the $f$-functions. In the rest of the paper, we
therefore concentrate on that formalism.

\subsection{Transport}
\label{sec:Transport}

In this section, we discuss the kinetic equation approach to transport. Our work program includes three principal steps: (i) we first compute the distribution functions describing the charge carrier populations at the nodes. Then (ii) we relate these distribution functions to charges and currents, before (iii) we extract observables from those. In the following, we discuss  this program on the concrete example of the transport problem sketched in Fig.~\ref{fig:1DCMESchematic}: we imagine the system biased by an external voltage $V$ in the $z$-direction, assumed to be colinear to an external field $B$. Our goal is to compute the conductance and its noise characteristic for a system whose dimensions are specified in the figure. 

\subsubsection{Distribution functions}

To obtain the distribution functions, we substitute the minimal ansatz $Q=\Lambda$
(cf. Eq.~\eqref{eq:QDist}) into the kinetic equation~\eqref{eq:Usadel_K}. For the
moment, we do not need to consider time dependent solutions, i.e. the distribution
function $f(\epsilon)$ becomes a function of a single energy argument, and
$[\partial_t,\Lambda]=0$. Assuming constancy of the distribution functions in directions perpendicular to $z$. Substitution of $\Lambda=\Lambda(f(\epsilon) )$, then  readily yields the
equations
\begin{equation}
\label{eq:KE}
 D\partial_z^2 f_n(\epsilon) +(-)^n \Gamma \partial_z f_n(\epsilon) -(-)^n \frac{1}{2\tau_n}
\Bigl( f_1(\epsilon) - f_2(\epsilon) \Bigr) = 0, 
\end{equation}
which have to be supplemented by the boundary conditions at the Ohmic contacts,
\begin{align}
f_n(\epsilon)\bigl|_{z=0} = f_F(\epsilon-V), \quad 
f_n(\epsilon)\bigl|_{z=L} = f_F(\epsilon).
\end{align}
The structure of these equations immediately reveals the presence of two distinct length scales:
first, equating diffusive and drift terms, $ D/l_m^2 \sim \Gamma/l_m$, we obtain the drift-diffusion crossover length
\begin{equation}
l_m = \frac{D}{\Gamma}= 4\pi^2 \frac{\sigma_{xx}}{B},
\end{equation}
which is related to the  magnetic length $l_B = (1/B)^{1/2}$ via the estimate
$l_m \sim l (l_B k_F)^2$.
Here $l = v \tau$ is the intra-node mean free path and $k_F = \mu/v$ is the Fermi
momentum. The analysis of our paper is limited to the classically weak magnetic
fields satisfying $l_B k_F \gg 1$, which gives us $l_m \gg l$. Physically, the scale $l_m$ discriminates between a short distance diffusive and a large distance drift transport regime. 
Second, the inter-node scattering becomes important on the length scale
\begin{equation}
l_n = \sqrt{ D \tau_n}.
\end{equation}
Owing to our assumption of weak inter-node scattering, $\tau_n/\tau \gg 1$, the latter length exceeds the intra node mean free path by far, 
$l_n \gg l$. However, the two mesoscopic length scales $l_m$ and $l_n$ can assume arbitrary relative values and this leads to the emergence of different types of transport. 

However, before discussing concrete solutions of the equations determining $f$, let us turn to the second step of the program, the determination of 

\subsubsection{Charge densities and currents}
\label{sec:hydrodynamics}

The discussion of this section applies independently to the fields of the two nodes, and we suppress the nodal index for brevity.
To relate the charge density $\rho(\mathbf{r},t) $ and current density
$\mathbf{j}(\mathbf{r},t)$ to the distribution function $h=1-2f$ in a general setting
with time dependence, it is convenient to express the latter in a Wigner
representation $h(t,\epsilon)$ defined through
\begin{align}
h(t_1, t_2) &= \int \frac{d\epsilon}{2\pi} h(t,\epsilon)e^{-i \epsilon (t_1-t_2)}, \cr 
&t = \frac{1}{2}(t_1 + t_2).
\end{align}
The charge density corresponding to $h$ is then given by
\begin{eqnarray}
\label{eq:rho}
\rho({\bf r},t) &=& -\frac{ \pi \nu}{2} {\rm tr}\, Q({\bf r},t,t)\tau_1 \\
&\stackrel{Q=\Lambda}{=}& -  \pi \nu \,h({\bf r}, t, 0) = - \frac{ \nu}{2} \int h({\bf r}, t,\epsilon) d\epsilon, \nonumber
\end{eqnarray}
where the matrix $\tau_1$ operates in the Keldysh basis. As one would expect, the
charge density is obtained by integration of the distribution function over energy
(we here assume constancy of the density of states over the energy windows relevant
to the problem.)

In the same manner, the current is obtained as 
\begin{align}
    \label{eq:CurrentDist}
    {\bf j}({\bf r},t) =\frac{1}{2}{\rm tr}\, ({\bf J}({\bf r},t,t)\tau_1),
\end{align}
where the \emph{matrix} current
$\mathbf{J}=\mathbf{J}_\mathrm{D}+\mathbf{J}_\mathrm{CS}$ is the sum of the diffusive
and the Chern-Simons current defined in Eqs.~\eqref{eq:D_current}
and~\eqref{eq:CS_current}. The evaluation of the former under the most general circumstances where an electric field is present requires a bit of care and is detailed in Appendix~\ref{app:CurrentDist}. As a result we obtain the components of the diffusive current as
\begin{eqnarray}
\label{eq:jDdist}
j_{i,\rm D}({\bf r},t) 
&=& - D \partial_{i} \rho ({\bf r},t) + \sigma_{xx} E_i({\bf r},t).
\end{eqnarray}
Here $\sigma_{xx}$ as before denotes the conductivity per node.
The two terms above are, respectively, Fick's and  Ohm's law. 
In a similar manner we obtain the Chern-Simons current 
\begin{eqnarray}
\label{eq:jCSdist}
 j_{i,\rm CS}({\bf r},t) &=& \pm  \frac{ \epsilon_{ijk} f_{jk}}{8\pi^2 \nu } \rho({\bf r},t),
\end{eqnarray}
where we neglected the topological winding number contribution in~\eqref{eq:CS_current}. In the case at hand, a magnetic field in $z$-direction, $\epsilon^{zjk} f_{jk} = 2  B$, the CS current is colinear to the field with
\begin{align}
    j_{z,\mathrm{CS}}= \pm \Gamma  \rho({\bf r},t),
\end{align}
and expresses the chiral magnetic effect (CME)~\cite{Alekseev:1998, Fukushima:2008}.
The sum
\begin{align}
    {\bf j}={\bf j}_\mathrm{D}+{\bf j}_\mathrm{CS},
\end{align}
is the full current density in the system.

We finally note that the charge and current density as derived above satisfy the continuity equation
\begin{equation}
\label{eq:NodalContinuity}
\partial_t \rho + \partial_{\bf r} {\bf j} =  \pm \frac{1}{4\pi^2 } ({\bf B}\cdot {\bf E} ),
 \end{equation}
where the term inhomogeneity on the r.h.s. is manifestation of the anomaly, and the sign change refers to the nodes. This equation, too, is proven in Appendix~\ref{app:CurrentDist}.

\subsubsection{Conductance \& shot noise}

\label{sec:MC_noise}

In the absence of a bulk electric field, the total current  in $z$-direction $I= \mathcal{A}j_z=\mathcal{A}(j_{z,1}+j_{z,2})$ as derived above is given by
\begin{align}
\label{eq:IQuasiOne}
I&=\mathcal{A}\sum_n\left(-D\partial_z \rho_n - \Gamma (-)^n\rho_n\right )=\cr
&=\mathcal{A}\nu \sum_n \int d \epsilon \left( - D \partial_z - \Gamma (-)^n \right )f_n(\epsilon).
\end{align}
Comparison with Eq.~\eqref{eq:KE} shows the constancy of the current $\partial_z
I=0$, as required for a static problem.

\begin{figure}[t]
\includegraphics[width=7.0cm]{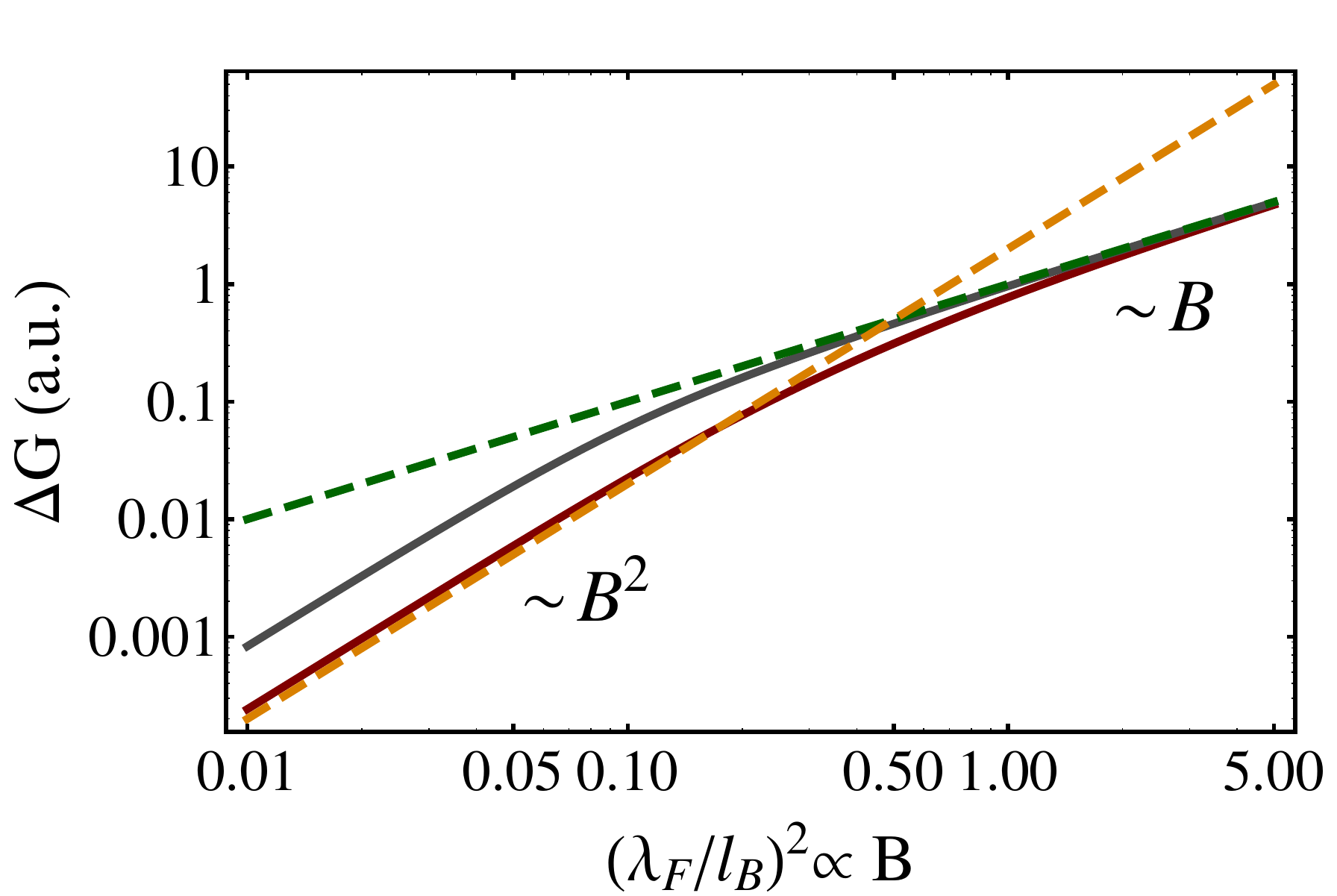}
\caption{Magnetoconductance $\Delta G(B) = G(B)- G(0)$ at a fixed length $L/l = 10^2$.
The bottom (red) line correspond to a wire with $L=10 l_n$ for which the nodes are
coupled by scattering. The upper (gray) line corresponds to a short wire $L/l_n \ll
1$ with effectively uncoupled nodes. Both curves demonstrate diffusion ($\propto
B^2$) to drift ($\propto B$) crossover in $G(B)$ upon increasing $B$.}
\label{fig:G_B}
\end{figure}

\emph{Conductance ---.} The next step is the solution of the second order linear differential equation~\eqref{eq:KE} subject to the voltage dependent boundary equation. The analytical solution of these equations is possible but rather cumbersome, and we do not display it here. (Readers interested in seeing the solution are advised to feed the equation into a computer algebra system.) Substitution of the solution into~\eqref{eq:IQuasiOne} and 
integration over the energy $\epsilon$ yields linear in $V$ electric current, $I = G(B) V$, 
where the conductance is given by  
\begin{eqnarray}
\label{eq:G_B}
G(B) &=&   \frac{N_\phi}{2\pi} \times g\left(\frac{L}{l_m},  \frac{L}{l_n} \right), \\
g(x,y)&=&\frac{ \left({y^2}/{x^2}+1\right)^{3/2}}{\frac{y^2}{2x}{\sqrt{{y^2}/{x^2}+1}}+\tanh \left(\frac{1}{2} x
   \sqrt{{y^2}/{x^2}+1}\right)}. \nonumber
\end{eqnarray}
Here $N_\phi = \Phi/\Phi_0$ denotes the number of flux quanta through the cross
section ${\cal A}$ as before, and the magnetic field enters through the dimensionless
parameter $x=L/l_m\propto B$. In the limit of zero magnetic field, $G(0) = 2 \nu D
/L$ matches the Drude result. However, in the limit of strong fields ($x \to \infty$)
we find $G(B) =   \frac{N_\phi}{2 \pi} \propto B$, i.e. $G(B)$ always becomes linear
in $B$ and independent of $L$ at high magnetic fields (cf. Fig.~\ref{fig:G_B}.)

The magnetoconductance shows its most interesting behavior in the regime $l_m \ll l_n$, i.e. when the drift diffusion scale is reached before the nodes get coupled by impurity scattering. In this case we have $ x \gg y$ and the 
crossover function $g(x,y)$ simplifies to
\begin{equation}
g(x,y) \simeq \frac{1}{{y^2}/{(2x)}+\tanh (x/2)}.
\end{equation}
Under these conditions, we may discriminate between long wires, $L > l_n$, for which
the nodes are effectively coupled by inter-node scattering, and shorter wires, for
which they remain uncoupled, $L < l_n$. For brevity, we will refer to the two types
of systems as `coupled' and `un-coupled' wires throughout.

\begin{figure}[t]
\includegraphics[width=7.0cm]{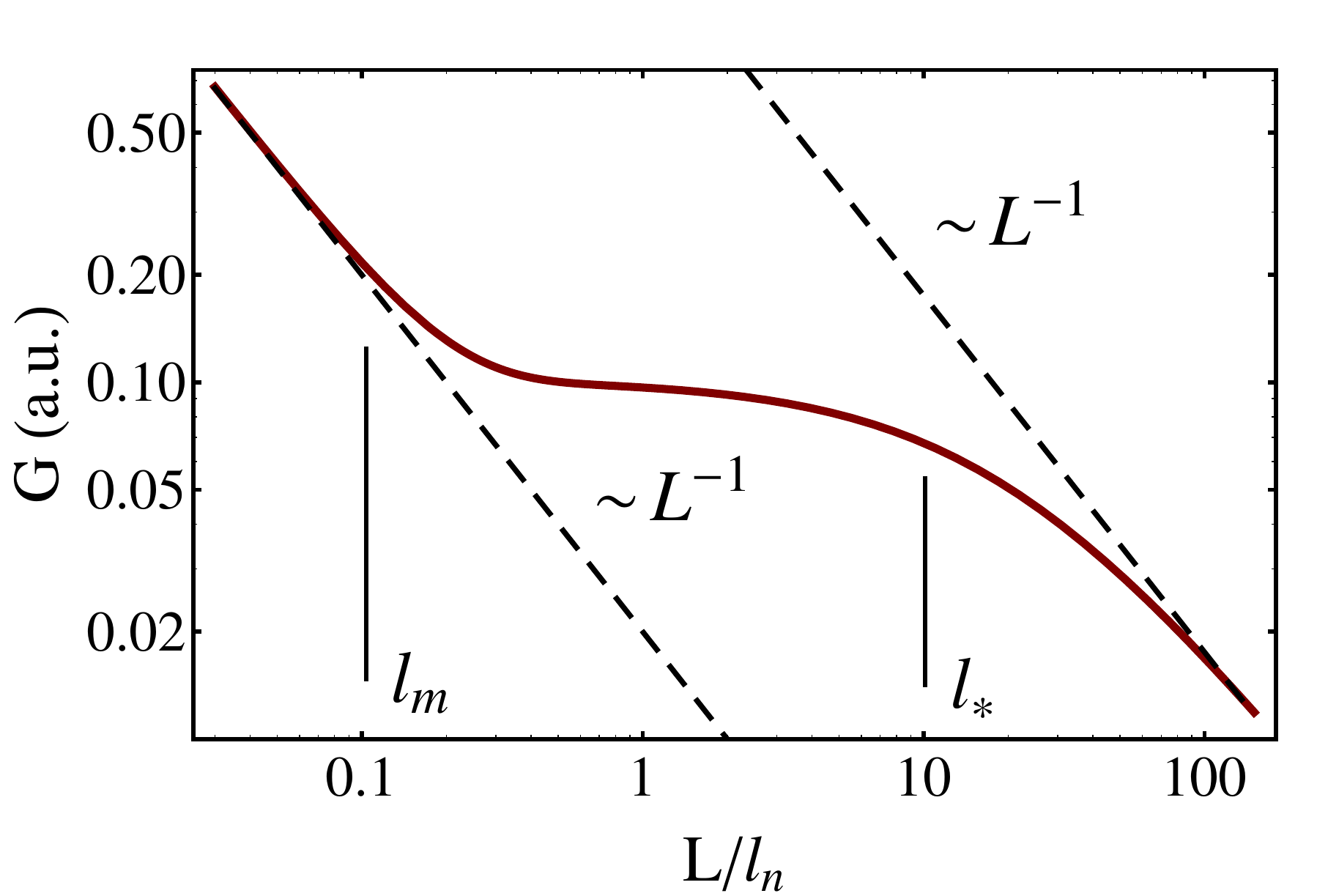}
\caption{Magnetoconductance versus the rescaled length $L/l_n$ shown for the fixed 
ratio of lengths $l_m/l_n = 0.1$ and the mean free path $l = 0.1 l_m$. }
\label{fig:G_L}
\end{figure}

For uncoupled wires, $L<l_n$, the magnetoconductance becomes
\begin{equation}
\label{eq:short_G}
G(B) \simeq  \frac{N_\phi}{2\pi} \coth(L/2 l_m), \quad (L,l_m) < l_n.
\end{equation}
As the function of length $L$ it shows diffusion ($\propto B^2$) to drift ($\propto
B$) crossover at  length scales $L \sim l_m$.  In the diffusive regime of short
lengths, $L < l_m$, the conductance shows Ohmic scaling, $G\sim L^{-1}$. However, for
larger lengths, $L>l_m$ a crossover into a `ballistic' drift dynamics takes place,
and the conductance saturates at the constant value $G\simeq N_\phi/2\pi$ (see Fig.~\ref{fig:G_L}.)

One might suspect that increasing the wire length to values, $L>l_n$, where the nodes get effectively coupled, lets the system re-enter a conventional diffusive phase. However, this is not so, up to the much larger crossover scale
\begin{align*}
l_\ast \equiv \frac{l_n^2}{l_m}>l_n,
\end{align*}
the conductance remains ballistic 
\begin{equation}
G(B) \simeq \frac{N_\phi}{2\pi}, \quad L<l_\ast. 
\end{equation}
The existence of this length scale can be rationalized from the structure of the differential equation Eq.~\eqref{eq:KE}, which tells us that up to lengths $L < l_\ast$ corresponding to
$\Gamma/L>\tau_n^{-1}$ the drift term overpowers the scattering term. 

For large lengths $L > l_\ast$ (or at weaker magnetic fields), the system re-enters a regime of diffusive dynamics, with Ohmic conductance
\begin{equation}
G(B) = \frac{N_\phi}{2\pi}  \frac{2l_n^2}{L l_m} = \frac{\mathcal{A}}{L} \frac{\tau_n}{2\nu} \left(\frac{B}{2\pi^2} \right)^2 , \quad L>l_\ast.
\end{equation}
Comparison with Eq.~(\ref{eq:sigma_B}) shows that this corresponds to the high field asymptotics of the earlier diagrammatic result. We encounter the high-field limit, $ \frac{\tau_n}{2\nu} \left(\frac{B}{2\pi^2} \right)^2 \gg \sigma_{xx}$ because we are working under the assumption $1 \ll l_m/l_n$ which is equivalent to the largeness of the magnetic field term compared to the bare conductivity. Fig.~\ref{fig:G_L} shows the conductance as a function of length. At intermediate lengths, $l_m < L < l_\ast$, we observe the formation of the ballistic plateau mentioned above. 

Finally, for weak magnetic fields such that $l_m \gg l_n$ no such behavior is found. In this regime, Eq.~\eqref{eq:G_B} reduces to Eq.~\eqref{eq:sigma_B}, where the field dependent correction now is weak compared to the Drude term.

\begin{figure}[b]
\includegraphics[width=7.0cm]{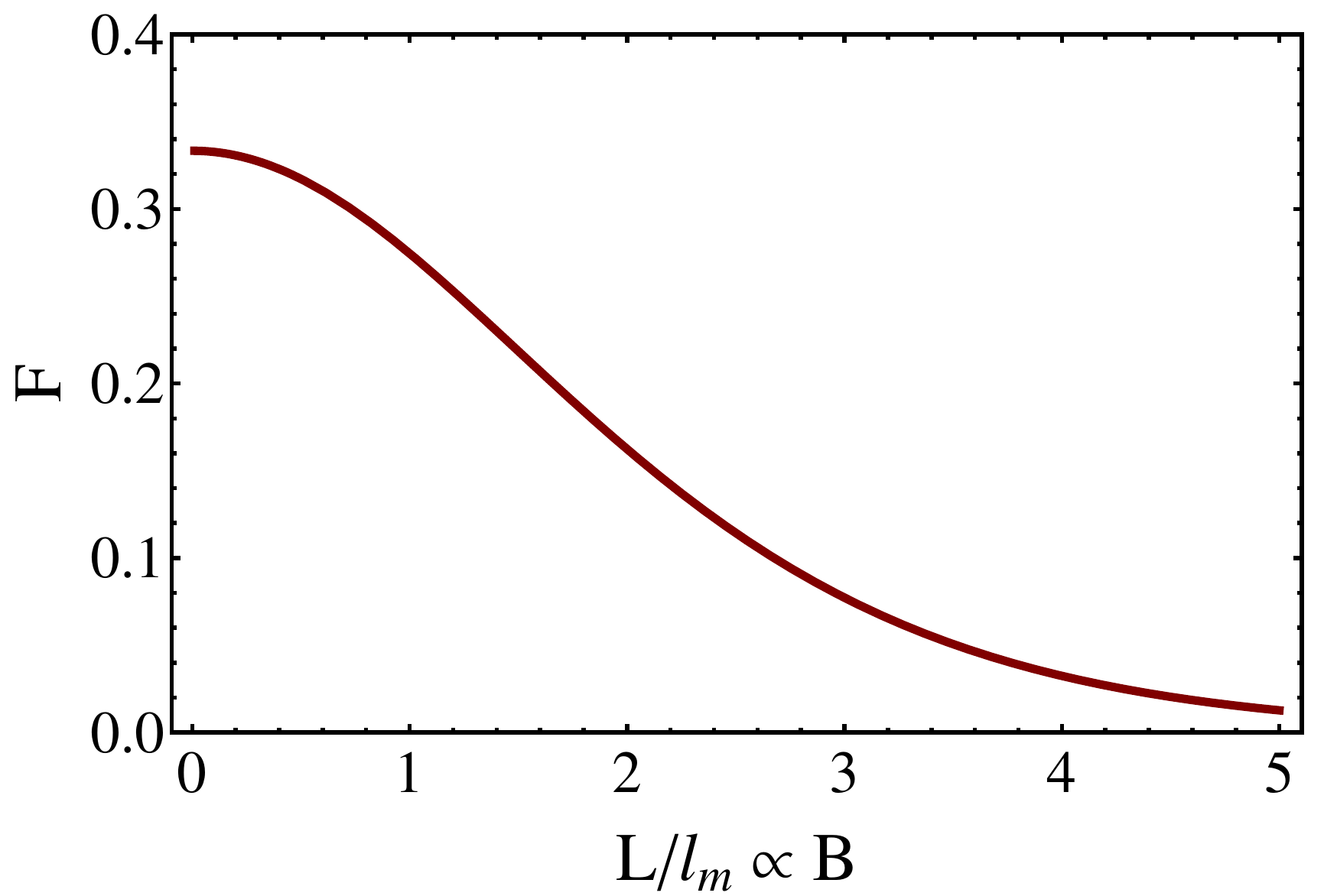}
\caption{Fano factor of the shot noise in for disordered wires shorter than the nodal equilibration length.
}
\label{fig:F_B}
\end{figure}

\emph{Non-equilibrium noise ---.} If the  drift transport regime identified above
really were governed by ballistic dynamics, then its noise characteristics should
differ from that of the diffusive regimes. In the following, we show that this is
indeed the case. Our object of study here is the field dependent Fano factor $F(B)$
of the current shot noise, i.e. the ratio of the observed shot noise to that of a
Poissonian process.   We will see below that the drift-diffusion crossover shows  in
the function $F(B)$ as a saturation to a value 1/3 at low $B$ ('diffusion') and an
exponential suppression at large $B$ (noiseless ballistic dynamics). The full profile
is shown in Fig.~\ref{fig:F_B}.

To simplify matters, we limit our consideration to short wires ($L \ll l_m$) for
which inter-node scattering can be neglected. In this case the channels of different
nodes do not mutually equilibrate, and one may expect deviation from diffusive noise
factor $F=1/3$. We compute the  shot noise within the framework of the theory of full
current statistics (FCS)~\cite{Nazarov:book}. Our starting point is the kinetic
equation (we here consider node no. 1, the equation for the other node differs in the
sign of the drift term; nodal index suppressed for brevity.)
\begin{equation}
\label{eq:Usadel_noise}
D \partial_z( Q \partial_z Q) -  \Gamma\, \partial_z Q = 0.
\end{equation} 
with   boundary conditions $Q\bigl|_{x=0,L} = Q_{L,R}$ modified to include sources
\begin{equation}
Q\bigl|_{L,R} = e^{i \frac{\chi_{L,R}}{2}\tau_1} \Lambda(f_{L,R})e^{-i \frac{\chi_{L,R}}{2}\tau_1}, 
\end{equation}
where $f_L = f_F(\epsilon-V)$ and  $f_R = f_F(\epsilon)$ as before, the Pauli matrix acts in Keldysh space, and $\chi_{L,R}$
are two auxiliary parameters, known as ‘counting fields’ in the theory of FCS. To
solve the kinetic equation we choose the similar parametrization for the $Q$-matrix
at any spatial point, 
\begin{equation}
\label{eq:Q_flambda}
Q(x) =  e^{ \frac{\lambda}{2}\tau_1} \Lambda(f)e^{- \frac{\lambda}{2}\tau_1}.
\end{equation}
Substituting Eq.~\eqref{eq:Q_flambda} into~\eqref{eq:Usadel_noise}, we find that the latter becomes equivalent to the set of two coupled nonlinear differential equations (cf. Appendix~\ref{app:FCS_action})

\begin{align}
\label{eq:f_n}
&   f'' + 2\,  \bigl( f (1-f)\lambda' \bigr)' -   f' = 0, \\
\label{eq:lambda_n}
&   \lambda'' -  (1-2f)\lambda'^{\,2} +  \lambda' = 0, 
\end{align}
where $f(x)$ and $\lambda(x)$ are expressed as functions of the dimensionless
parameter $x=z/l_m$ used previously. Which have to be supplemented with the boundary
conditions $f(x=0,L/l_m)=f_{L,R}$ and $\lambda(x=0,L/l_m)=i\chi_{L,R}$. Focusing on
zero temperature and the ‘transport energy’ regime $0 < \epsilon < V$, they simplify
to $f(0)=1, f(L)=0$. We now use the fact that the kinetic
equations~(\ref{eq:Usadel_noise}) conserve the matrix currents $J_n = D \partial_z Q + (-)^n\Gamma Q$,
where we re-introduced the nodal index. 
This suggests to define the likewise conserved full scalar  current $J_z(\chi) =
\sum_n {\rm tr}(J_n \tau_3)$. Within our
choice of parametrization it takes the form
\begin{eqnarray}
\label{eq:Jz}
J_z(\chi) 
&=& \sigma_{xx}\mathcal{A}\sum_n \int d\epsilon \Bigl[ -  f_n' - 2  f_n(1-f_n) \lambda_n'  \nonumber \\
&-&(-1)^n  f_n \Bigr].
\end{eqnarray}
The scalar current  depends only on the phase difference $\chi = \chi_L - \chi_R$. According to the general principles of FCS, it yields the physical current and the current noise as 
\begin{align*}
I=  J_z|_{\chi=0},\qquad S = i  (\partial_\chi J_z)|_{\chi=0},
\end{align*}
respectively. The noise may therefore be obtained by first oder perturbative solution of the differential equations~\eqref{eq:f_n} in $\chi$. To zeroth order, the first of the equations simplifies to  
$({f^0_n}{'} + (-1)^n  {f^0_n})' = 0$, while the second becomes $\lambda{''}+\lambda'=0$. Expressed in terms of the solutions of these equations, the equation for the first order correction $f^1$ assumes the form  
\begin{equation}
{f^1_n}''  + (-1)^n  {f^1_n}' = - 2\, \bigl( {f^0_n} (1-{f^0_n})\lambda_n' \bigr)'.
\end{equation}
The solution of these equations is straightforward, and upon substitution of the result into~\eqref{eq:Jz}, we obtain the conductance in agreement with our previous discussion. For the noise we obtain $S=FI$, where the Fano factor is given by (cf. Fig.~\ref{fig:F_B})
\begin{eqnarray}
F(x) &=& \frac{\sinh x - x}{2 \sinh^2(x/2) \sinh x} \nonumber \\
&=&
\left\{ 
\begin{array}{cl}
{1}/{3}- {x^2}/{15} + \dots, & \quad x \ll 1 \\
2 e^{-x}, & \quad x \gg 1
\end{array} \right.,
\end{eqnarray}
where $x=L/l_m$. This equation tells us that the noise vanishes exponentially with
the magnetic field. This is because for increasing field, the system sustains a
larger number of effectively ballistic 1D transport channels, which superimpose to support a transport regime effectively void of scattering. For short systems $L<l_m$ the transport is diffusive and the noise function crosses over to the standard value $F=1/3$. Finally, we expect that for wires of length $L>l_\ast$, i.e. longer than the nodal equilibration length a crossover back to $F=1/3$ takes place, i.e. the noise function is governed by a pronounced inhomogeneity at intermediate length $l_m\lesssim L\lesssim l_\ast$.

\begin{figure}[t]
\centering{
\includegraphics[width=0.3\textwidth]{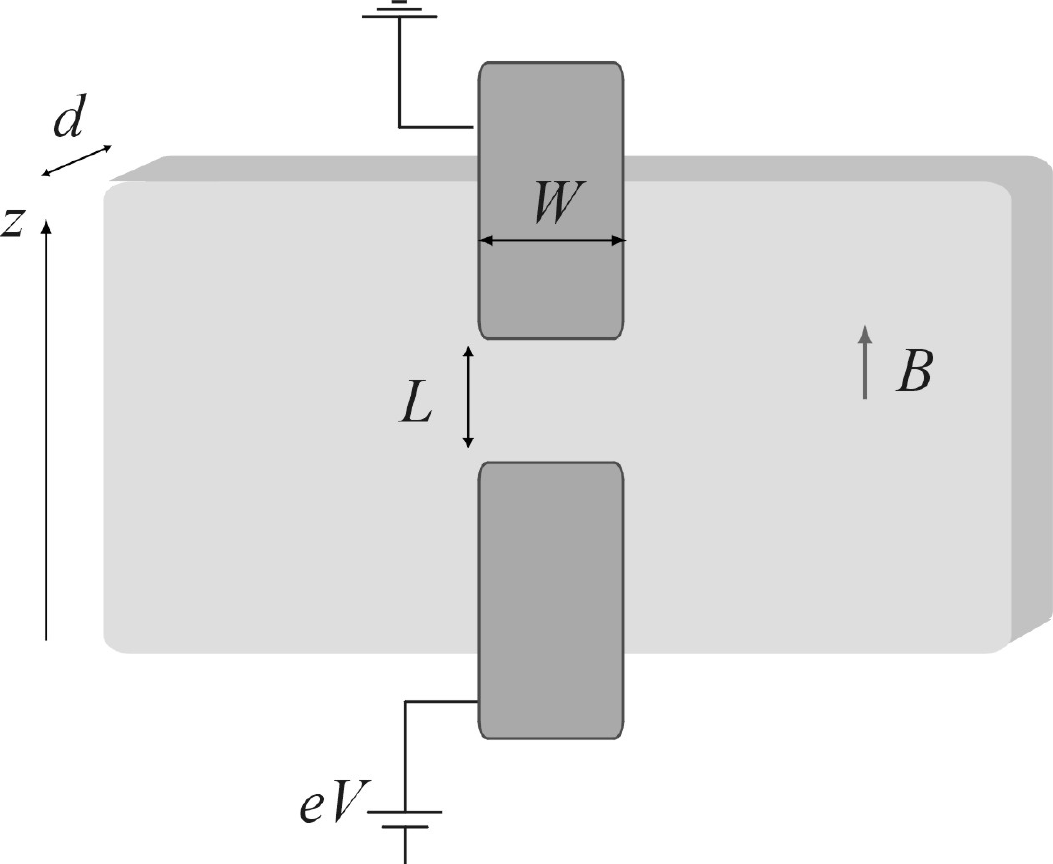}
}
\caption{The top view of the mesoscopic contact made of the Weyl semimetal film of thickness $d$ and 
two rectangular-shaped Ohmic contacts of width $W$ placed at distance $L$ on the top of the film. }
\label{fig:Device1}
\end{figure}

While our discussion above focused on a quasi one-dimensional idealization of the system for simplicity,  
we believe that qualitatively similar results hold for the 2D geometry shown in Fig.~\ref{fig:Device1}.
For this setup the amount of 1D channels effectively contributing to the conduction will be given by the number of flux quanta through the cross section of the contact, $N_\Phi = A B_z /\Phi_0$ with  $A \sim W d$. 

\subsubsection{Quasi one-dimensional localization}

\label{sec:1dloc}

An interesting question to ask is whether the drift dynamics at intermediate scales will overpower Anderson localization effects which are known to be strong in one dimension. Concerning this question, nothing can be learned from the kinetic equation approach, which does not take the quantum fluctuations driving localization into account. Rather, we need to get back to the field theory formalism of section~\ref{sec:GFT}. Compared to the general discussion of that section, we now consider a simpler situation, where the geometry is confined with fluctuation fields $T(z)$ varying only along the axis of a wire, and the external vector potential $a_i= -(B/2)\epsilon^{3ij}x_j$ generating a magnetic field along the wire axis.

Under these circumstances, the effective action of the system reduces to
\begin{align}
\label{eq:SeffQuasi1}
&S[T]= \sum_n S_n[T_n]-\frac{\pi\nu\mathcal A}{4\tau_n}\int dz\, \mathrm{tr}(Q_1 Q_2),\\
&\;S_n[T]=\int dz\,\mathrm{tr}\left(\frac{\sigma_{xx} \mathcal{A}}{8} \partial Q \partial Q
+\frac{(-1)^n N_\phi}{2}\,T^{-1}\partial T \right),\nonumber
\end{align}  
with derivatives $\partial=\partial_z$. To understand the emergence of this structure
from the full  action, first note that the topological term $S_\mathrm{top}$ drops
out in the quasi one-dimensional limit, due to its mixed spatial derivatives. The
drift term linear in derivatives descends from the CS action, more precisely from the
contribution $S_{CS}^{(1)}$ to the expansion~\eqref{eq:CS1_Qa_app} of the latter in
the external vector potential. Noting that the combination
$\epsilon^{ijk}\mathrm{tr}(Q\partial_i Q\partial_j Q)=4
\epsilon^{ijk}\partial_i\mathrm{tr}(T^{-1}\partial_j T)$ can be written as a full
derivative, a straightforward integration by parts in $S_{CS}^{(1)}$ immediately
produces the drift term. Finally, we consider the action above as a replica action,
i.e. without time and distribution function dependence, as neither are relevant to our present discussion.

A variation of the $T$-fields in Eq.~\eqref{eq:SeffQuasi1} produces the kinetic
equation~\eqref{eq:KineticQuasi1} (without the time derivative term). In fact, the
action above could have been derived as the unique action producing the kinetic
equations from a variational principle. If it were not for the term linear in
derivatives, the action above would resemble that of either two uncoupled actions
$\sum_n S_\mathrm{diff}[Q_n]$ ($\lim \tau_n \to \infty$), or of a single action
$2\times S_\mathrm{diff}[Q]$  ($\lim \tau_n \to 0 \Rightarrow Q_1=Q_2=Q$), where
$S_\mathrm{diff}$ is the standard diffusive action (the first term in
Eq.~\eqref{eq:SeffQuasi1}). In either case, this action describes strong Anderson
localization at length scales exceeding the localization length $\xi \sim
\sigma_{xx}\mathcal{A}\sim N l $, where $N\sim k_F^2\mathcal{A}$ is the number of
transverse channels of the system. (To be more precise, the quantitative description
of Anderson localization requires the transcription of the action from a
Keldysh/replica framework to supersymmetry, more on this point below.) This prediction conforms with the qualitative expectation that the dimensional reduction of a Weyl system in diffusion dominated regimes should show Anderson localization. 

How will the drift term affect this picture? Once more, let us focus on the physics
of an isolated node first, $\tau_n\to \infty$, as described by a single copy
$S_n[T]$. Dimensional analysis suggests that at large length scales the term linear
in derivatives will dominate over the two derivative term. The linear contribution is
known as the action of the 'ballistic sigma model'. It appears, e.g. in the context
of clean chaotic quantum systems~\cite{Muzykantskii:1995, Andreev:1996}, 
or as an effective description of quantum Hall edges~\cite{Pruisken:1999}, 
and in either case describes directed motion unaffected by scattering. The
dimensional argument suggests that this type of dynamics prevails in the present
context, provided the drift term dominates over the diffusion term, i.e. at scales
$L\gtrsim l_m$, which is the regime identified above as the drift transport regime.
To substantiate that expectation, one would need to (a) reformulate the action in a
supersymmetry~\cite{Efetov1997Sypersymmetry} framework. In a manner we will not discuss in detail this would
change the ‘target space’ of the $T$-fields from the unitary group $\mathrm{U}(2R)$
to the super-group $\mathrm{U}(2|2)$, but leave the structure of the action
unchanged. In a second step, (b), one would then reformulate the one-dimensional
field integral in terms of an equivalent ‘transfer matrix’ equation (conceptually,
this is the `Schroedinger equation' of the path integral, not to be confused with
the variational ‘Euler-Lagrange equations’ discussed above), and solve the latter. In this language, localization corresponds to the appearance of a finite gap $\Delta \equiv \xi^{-1}$ in the spectrum of the corresponding Hamilton operator, and length corresponds to ‘time’, $t\equiv L$ such that $\exp(-\Delta t)=\exp(-L/\xi)$ describes a decaying/localizing long time/distance behavior, if the gap is finite.  
The diffusive term leads to a transfer matrix operator describing free motion on the  target
space, $\mathrm{U}(2|2)$, and the corresponding, gapped, free particle spectrum leads to localization.   The linear term couples the free motion to an fictitious field. A sufficiently strong perturbation of this type is expected to remove the spectral gap and hence to render the long distance transport ballistic. Eventually, at length scales $L>l_\ast$, the node mixing term exceeds the drift term in strength, and this will lead back to a conventional localization picture on the field space $T_1=T_2=T$ of the coupled nodes.  However, a quantitative formulation of that mechanism is beyond the framework of this paper.

\section{Conclusions}
\label{sec:Conclusions}

In this paper, we have developed and discussed the effective theory describing the
disordered Weyl (semi)metal. We have found that that theory resembles that of a three
dimensional Anderson metal, with the important modifications that two topological
terms significantly modify the transport physics of the system. These two terms are,
respectively, a three dimensional ‘layered’ extension of the $\theta$-term crucial to
the description of the quantum Hall effect, and a Chern-Simons term. Loosely speaking
they describe the various manifestations of the anomalous Hall effect and of the
chiral magnetic effect, respectively.

In the presence of the $\theta$-term reflects the conceptual similarity of the Weyl
metal to a layered system of two-dimensional topological insulators. Much as with
ordinary quantum Hall insulators, the coupling of the two-dimensional compounds  is
expected to generate a metallic layered quantum Hall phase, distinguished from an
ordinary three dimensional metal by the presence of a stable Hall conductivity.
Building on earlier field theoretical analyses of the layered quantum Hall system, we
could quantitatively corroborate the expected universality of the Hall conductivity
at arbitrary length scales.

With the Chern-Simons theory, the situation is more involved in that the integrity of
the latter depends on the coupling between the two Weyl nodes. In the limit of strong
scattering between the  nodes the  Chern-Simons actions mutually cancel out and we
are left with the situation outlined above. Physically, this means a restoration of
parity invariance via the hybridization between parity distinct nodes. However,
parametric differences in the intra- and inter-node scattering rates are a very
concrete option, and this opens the room for the existence of extended crossover
regimes, in which the Chern-Simons theory is phenomenologically visible. These are
the regimes in which the chiral magnetic effect unfolds. Perhaps most interestingly,
we have found that the corresponding transport physics is one of drift diffusion
dynamics, where diffusion at short scales gives way to quasi-ballistic drift at
larger scales, before the node-coupling kicks in, and diffusion takes over again. In
the intermediate regime, the current flow is genuinely ballistic (noiseless), the
presence of disorder notwithstanding. We also argued that the drift transport is
effectively protected against Aderson localization (in confined geometries, where the
latter may be an issue.)

Extending earlier perturbative results, our analysis shows that the high sensitivity
of Dirac nodes to the presence of disorder is not in contradiction to the formation
of highly universal large distance transport regimes. The topology of the nodes
(which is encoded in their Chern numbers) survives in topological terms of the field
theory, which in turn affect transport coefficients. It will be interesting to
explore how such protection generalizes to other system classes, notably those
supporting topological line effects instead of isolated points.

\section{Acknowledgments}  

Discussions with V.~Gurarie, P.~Brouwer and A.~Stern are gratefully acknowledged.

\appendix

\section{Fast momentum integrals} 
\label{sec:fast_momentum_integrals}
In this appendix, we provide some details on the evaluation of the fast momentum integrals appearing in the main text. Straightforward matrix algebra brings the integral $f_{ss'}$ of Eq.~\eqref{eq:f_ss} into the form 
\begin{eqnarray}
\label{eq:f_ss_explicit}
f_{ss'} 
 & = & \int dp\, \left(-\frac23 p^2 + 2\epsilon_{s} \epsilon_{s'}\right) N_p^s N_p^{s'}. 
\end{eqnarray}

Let us  first consider the integral $f_{ss}$. 
The momentum integral in the first term of Eq.~(\ref{eq:f_ss_explicit}), $\sim p^2$, requires regularization,
\begin{align}
& \frac{2}{3}\int dp\, p^2 (N_p^s)^2  
\overset{\rm reg.}{\to} \frac{2}{3}\int_0^{+\infty} \frac{dp}{2\pi^2} \left[ \frac{p^2}{(p^2 - \epsilon_s^2)^2} - \frac{p^2}{(p^2 + \eta^2)^2}\right] \nonumber \\
& \overset{\eta \to 0}{=}  \frac{\epsilon_s^2}{3\pi^2} \int_0^{+\infty} dp\, \frac{2p^2 - \epsilon_s^2}{(p^2 - \epsilon_s^2)^2} = 
\frac{i s \epsilon_s}{4\pi}.
\end{align}
The regulator above follows from the formal expansion of the action $S_{\eta}[A]$, cf. Eq.~(\ref{eq:action_reg}).
Turning to the term $\sim \epsilon_s^2$ in Eq.~(\ref{eq:f_ss_explicit}), we find $2\epsilon_s^2\int d p (N_p^s)^2  = i s \epsilon_s/ 4\pi$, which is UV finite. Combing two contributions one arrives at $f_{ss} = 0$.

Let us turn to the evaluation of $f_{s,-s}$. As before, the first contribution, $\sim p^2$, has to be regularized
and we obtain
\begin{align}
& -\frac{2}{3}\int dp\, p^2 N_p^+ N_p^- 
\overset{\rm reg.}{\to} -\frac{2}{3}\int_0^{+\infty} \frac{dp}{2\pi^2} 
\Biggl[ \frac{p^2}{(p^2 - \epsilon_-^2)(p^2 - \epsilon_+^2)}  \nonumber \\
& -  \frac{p^2}{(p^2 + \eta^2)^2}\Biggr]\,
\overset{\eta \to 0}{=} 
\,\frac{3\kappa^2 - \epsilon^2}{12 \pi \kappa}.
\end{align}
As to the second contribution, one finds 
\begin{equation}
2 \epsilon_+ \epsilon_- \int d p\,(N_p^+ N_p^-) = {(\epsilon^2 + \kappa^2)}/4\pi\kappa. 
\end{equation}
The sum of the latter two terms then yields $f_{s,-s} = \sigma_{xx}^{1}$ as given in Eq.~\eqref{eq:f_ss_res}. 

We next turn to the ‘anisotropic version’ of the above integral, Eq.~\eqref{eq:fssprimeDef}, relevant to the computation of the Hall coefficients. Doing the straightforward trace over Pauli matrices, we find
\begin{align}
f^{i\neq j}_{ss} &= 0,  \nonumber \\
\label{eq:f_xy_pm}
f^{i\neq j}_{s, -s} &= 4 s \epsilon^{3 i j} \kappa \int dp\, (p_3 + b) N_p^+ N_p^-. 
\end{align}
The remaining integral can be evaluated in cylindrical coordinates with the measure 
$d p = p_{\perp }\, d p_{\perp} dp_3  /4\pi^2$. First integrating over the  
lateral momentum $p_{\perp }$, we obtain
\begin{eqnarray}
f^{12}_{+,-} &=& \frac{i}{8\pi^2\epsilon}\int_{-\Lambda}^\Lambda d p_3\, (p_3 + b) 
\ln\left[\frac{\epsilon_+^2  - (p_3 + b)^2 }{\epsilon_-^2  - (p_3 + b)^2 } \right] \nonumber \\
&=& \frac{2 \kappa }{\pi^2} \left( \frac{b}{\Lambda}\right) + {\cal O}(\Lambda^{-2}),
\end{eqnarray}
as stated in the text.

\section{Gauge field regulator action} 
\label{sec:RegulatorField}

In this appendix, we discuss the non-trivial dependence of the regulator action ${\cal S}_\eta[\bar A]$ on an external gauge field.   
We know from Redlich's paper~\cite{Redlich1984Parity} that this dependence can be summarized in three points:
\begin{itemize}
\item[(i)] the regulator changes under gauge transformations by $i\pi (n_+ -n_-)$ where $n_\pm$ are winding numbers of $k$ associated with
homotopically non-trivial {\it right} gauge transformations introduced as $T' = T k$ --- 
it also follows from definition~(\ref{eq:A_bar}) that $\bar A$ transforms in this case as 
$\bar A'_i = k^{-1} A_i k + k^{-1}\partial_i k$, which is compatible with the transformation 
law discussed in Sec.~\ref{sec:Reg};
\item[(ii)] the regulator serves as a UV counter-term to the first part of the action,  ${\cal S}[A,a]$.
\item[(iii)] In the presence of a non-vanishing field-strength tensor 
the finite part of the regulator introduces contributions which are non-analytic
functions of $\bar F_{jk}$. 
\end{itemize}
This latter point did not play a role  in the discussion of the field free action, $a=0$,  since the field $A_i=T^{-1}\partial_i T$ by itself was a full gauge and
had zero field-strength tensor. However, in the presence of an external field, we obtain the non-vanishing~\eqref{eq:F_jk}, which is gauge equivalent to the external tensor $f_{jk}$ in~\eqref{eq:f_jk}. It is hard to make general statements regarding the ensuing terms for completely arbitrary $f$. However, as stated in the main text, we may assume triviality of the external field in both  retarded-advanced and replica space, i.e. 
$f_{jk} \propto \mathds{1}_{ra} \otimes \mathds{1}_R$. Under these conditions,  the non-analytic part of ${\cal S}_\eta$ becomes proportional
to the number of replicas and thus vanishes in the limit $R \to 0$.

\section{Derivation of the gauged effective action}
\label{sec:DerivationGaugedAction}

In this appendix, we briefly discuss the derivation of the gauged
action~(\ref{eq:action_Ta}).  To obtain the diffusive action $S_d$ one should
reiterate the steps of Sec.~\ref{sec:SdDer} which  led us to
Eq.~(\ref{eq:Id2}). At this point $A_i$ should be replaced by $\bar A_i$ and thereby
Eq.~(\ref{eq:Id2}) transforms into $\mathrm{tr}[\tau_3,\bar A_i]^2 =
\mathrm{tr}(\nabla_i Q)^2$, which finally yields the gauge invariant extension of the
diffusive term. The topological action is derived from the
representation~(\ref{eq:S_II_top_A}), as before. Changing $A_i \to \bar A_i$ and using 
Stoke's theorem one arrives at
\begin{equation}
\label{eq:S_top_contour}
    S^{II}_\mathrm{top}[A,a] = \frac{1}{2} \sigma_{xy}^{II} \sum_{i=1,2}
\oint_C \,\mathrm{str}(\tau_3 A_i - i Q a_i) dx^i dx_3
\end{equation}
It can be checked by the direct inspection that the difference of this action to the
one given by Eq.~(\ref{eq:Q_top}) evaluates to
\begin{equation}
S_\mathrm{top}[A,a]- S^{II}_\mathrm{top}[A,a] = \frac{i\sigma_{xy}^{II}}{4} \epsilon^{3ij} 
\int dx\, \mathrm{tr}( Q f_{ij}),
\end{equation}
which is zero if $f_{jk}$ satisfies  our assumption of proportionality to the
identity matrix, $f_{jk} \propto \mathds{1}_{ra} \otimes \mathds{1}_R$.

Finally, the derivation of the CS action $S_{\rm CS}[\bar A]$ closely parallels those
of Sec.~\ref{sec:SCSDer}. Collecting  terms of order ${\cal O}(\bar A^2 q)$ and
${\cal O}(\bar A^3)$ in the gradient expansion one arrives at the intermediate
result, cf. Eqs.~(\ref{eq:A2_1}) and (\ref{eq:A3_0}),
\begin{align}
\label{eq:SC_E}
{\cal S}^{(2)}_1[\bar A]+{\cal S}^{(3)}_0[\bar A] & = S_{CS}[\bar A]  \\ 
& - \epsilon^{ijk}\,\frac{\epsilon}{12\pi\kappa}  
\int dx \,\mathrm{tr}(\bar A_i P^+ \bar F_{jk} P^-). \nonumber
\end{align}
Taking into account that the field-strength tensor~(\ref{eq:F_jk}) simplifies to
$\bar F_{jk} = - i f_{jk}$ and therefore commutes with projectors, $[f_{jk}, P^\pm ] = 0$,
one observes that the energy term in the last expression vanishes. Hence, we can
conclude that Eq.~(\ref{eq:action_Ta}) is indeed the required gauge invariant
extension of our action.

In passing we note that in the case of the Weyl semimetal, $\epsilon \ll \kappa$,
the energy term in Eq.~(\ref{eq:SC_E}) can be disregarded. 
In this limit the action~(\ref{eq:action_Ta}) with the CS term $S_{\rm CS}[\bar A]$ 
describes the disordered Weyl semimetal 
for arbitrary non-Abelian gauge fields $a_i$, without any restrictions to $f_{jk}$.
Even though our practically inclined calculations below are performed for 
the Abelian physical electro-magnetic field, the non-Abelian gauge invariance of the action~(\ref{eq:action_Ta}) 
is an appealing and important ingredient of our field theory. 
It helps, for instance, to identify the equation of motion for the matrix field $Q$, as discussed in the next subsection.

\section{Derivation of the kinetic equation}
\label{app:Usadel}

The kinetic equations are obtained by linearization of the action~(\ref{eq:action_Ta}) in $\lambda$, using the
variations~(\ref{eq:variations}). Let us first set the external gauge fields $a_k$ to
zero. In this case it is technically more convenient to use the CS action in the
following form
\begin{align}
\label{eq:CS_simple}
S_\mathrm{CS}[A]=-\frac{i\epsilon^{ijk} }{16\pi}\int dx \,\mathrm{tr}\left( \partial_i A_{j}  \tau_3  A_{k}+\frac{1}{3} \tau_3 A_{i}   \tau_3 A_{j}  \tau_3 A_{k}\right),
\end{align}
which is a variant of the full action~(\ref{eq:SC_tau}) where we can use $\partial_i A_j = - A_i A_j$.
For the diffusive part one then obtains
\begin{equation}
\delta S_{\rm d}[Q] = \pi\nu \int dx \,{\rm tr}\, \lambda 
\Bigl( - D \partial_{\bf r} (Q \partial_{\bf r} Q) \Bigr) ,
\end{equation}
while the CS action gives
\begin{equation}
\label{eq:var_CS}
\delta S_\mathrm{CS}[A]= \frac{i}{16\pi}\int dx \,\epsilon^{ijk} \sum_{\alpha,\beta,\gamma}\mathrm{tr} \,
\lambda \, R^{ijk}_{\alpha\beta\gamma} ,
\end{equation}
where we have defined
\begin{eqnarray}
R_{111}^{ijk} &=& \partial_i T \tau_3 T^{-1} \, \partial_j T \tau_3 T^{-1}\, \partial_k T \tau_3 T^{-1}, \\
R_{211}^{ijk} &=& T \tau_3 \partial_i  T^{-1} \, \partial_j T \tau_3 T^{-1}\, \partial_k T \tau_3 T^{-1}, 
\, \mbox{etc.} \nonumber 
\end{eqnarray}
The Greek indices in $R^{ijk}_{\alpha\beta\gamma}$ take values $1$ and $2$ and
indicate if the derivative operator $\partial_i$ acts on $T$ or on $T^{-1}$, resp.
Upon summation over all Greek indices, one arrives at the CS contribution to the
kinetic equation.

In the general case of non-vanishing external gauge fields $a_k$, the CS action is
given by Eq.~(\ref{eq:SC_tau}) in Sec.~\ref{sec:Gauge_inv_action}, and depends on
$a_k$ through the gauge invariant vector potential $\bar A_k  = A_k   - i T^{-1} a_k
T$. For practical calculations it is advantageous to rewrite it in a form where the
gauge fields $a_k$ are singled out explicitly,
\begin{equation}
S_\mathrm{CS}[A, a] = S_\mathrm{CS}[A] + \sum_{l=1}^3 S^{(l)}_\mathrm{CS}[Q, a],
\end{equation}
where the functionals $S^{(l)}_\mathrm{CS}[Q, a]$ read
\begin{align}
\label{eq:CS1_Qa_app}
S^{(1)}_\mathrm{CS}[Q, a]  &=    - \frac{1}{16\pi}\int dx \,\epsilon^{ijk} \mathrm{tr}
\left(  Q \partial_i Q \partial_j Q a_k \right), \\ 
 S^{(2)}_\mathrm{CS}[Q, a]  &=  \frac{i}{16\pi}\int dx \,\epsilon^{ijk} \times \nonumber \\
& \qquad \mathrm{tr}\Bigl(   Q \partial_i Q a_j Q a_k  
-  a_i a_j \partial_k Q  + 2\, a_i (\partial_j a_k) Q \Bigr), \nonumber \\
S^{(3)}_\mathrm{CS}[Q, a]  &=  \nonumber
 \frac{1}{16\pi}\int dx \,\epsilon^{ijk} \mathrm{tr}
\Bigl(  a_i a_j Q a_k  + \frac{1}{3}\, a_i Q a_j Q a_k Q  \Bigr). \nonumber 
\end{align}
Although the form of $a$-dependent terms  is difficult to motivate, the result above
demonstrates the important structure of the theory --- $S_{\rm CS}^{(l)}[Q,a]$ are
functionals of the fields $Q$ and $a_k$ only and hence are explicitly invariant under
the {\it right} gauge transformation $T' = T k$. Therefore the arguments of
Sec.~\ref{sec:SCSDer} in which we discuss the right gauge invariance of our
$\sigma$-model action remain valid also in the presence of external gauge fields
$a_k$.

Going back to the derivation of the gauge invariant form of the kinetic equation 
we note that the most efficient way to proceed is not to vary the action
over $\delta Q$ directly but rather to go an indirect way and calculate the matrix 
current $J= J_D + J_{\rm CS}$, as  explained in Sec.~\ref{sec:Saddle_point}. 
Using the form of the CS action~(\ref{eq:CS1_Qa_app})
one then arrives at the expression~(\ref{eq:CS_current}) for the CS current 
and using the diffusive action $S_{\rm d}[Q,a]$~(\ref{eq:Q_diff}) one finds $J_D$. 
The  kinetic equation~(\ref{eq:Usadel_a}) then follows from the condition $\nabla_k J^k = 0$.

\section{Current density from distribution functions}

\label{app:CurrentDist}

In this appendix, we discuss some technical details relating to the representation of the current in terms of distribution functions. As discussed in the main text, we   split the current into a diffusive and a CS part, 
${\bf j}({\bf r},t) = {\bf j}_{\rm D}({\bf r},t) + {\bf j}_{\rm CS}({\bf r},t)$. 
From the diffusive part of the matrix current~(\ref{eq:D_current}),
\begin{equation}
{\bf J}_{\rm D}({\bf r},t_1, t_2) = \pi\nu D ( Q \nabla Q)_{t_1,t_2}
\end{equation}
the diffusive current follows as
\begin{equation}
{\bf j}_{\rm D}({\bf r},t) =\frac{1}{2}{\rm tr}\, ({\bf J}_D({\bf r},t,t)\tau_1),
\end{equation}
(we have temporary omitted the nodal index for brevity). This expression has to be
understood as the limit $t_+ \to t$ and $t_- \to 0$ in order to regularize the
covariant derivative $\nabla$ containing the vector potential. For any distribution
function $h$ one has the following universal short time asymptote
\begin{equation}
\lim_{\delta t \to 0} h(t, t+\delta t) = -\frac{1}{i\pi\delta t} + {\cal O}(1),
\end{equation} 
which follows from the fact that all electron states deep below the Fermi energy are filled while high energy states are empty. (In the definition for charge density~(\ref{eq:rho}) we assume that the singular part is subtracted,
the latter is equivalent to taking the principal value of energy integral at $\delta t=0$.)
This yields
\begin{eqnarray}
\label{eq:[aQ]}
-i [{\bf a}, Q]_{t,t+\delta t} &=& \left(
\begin{array}{cc}
0 & - 2i\, [{\bf a},h]_{t,t+\delta t} \\
0 & 0
\end{array}\right) \\
&\stackrel{\delta t\to 0}\longrightarrow& 
\left(
\begin{array}{cc}
0 & 2   {\bf E}(t)/\pi  \\
0 & 0
\end{array}\right), \nonumber
\end{eqnarray}
where we have used $\partial_t {\bf a} = -  {\bf E}$. Hence the diffusive current reads
\begin{eqnarray}
{\bf j}_{\rm D}({\bf r},t) &=& \pi \nu D \, \partial_{\bf r} h({\bf r},t,0) +  \nu D\, {\bf E}({\bf r},t)
\nonumber \\
&=& - D \partial_{\bf r} \rho ({\bf r},t) + \sigma_{xx} {\bf E}({\bf r},t),
\end{eqnarray}
whose component representation is given in Eq.~\eqref{eq:jDdist}.
 
To find the CS currents we use Eq.~(\ref{eq:CS_current}) to obtain
\begin{eqnarray}
 j^z_{\rm CS}({\bf r},t) &=& \frac{1}{2}{\rm tr}\, J^z_{\rm CS}({\bf r},t,t)\tau_1.
\end{eqnarray}
We substitute  $Q=\Lambda$ into (\ref{eq:CS_current}), ignore the winding number contribution, and immediately arrive at Eq.~\eqref{eq:jCSdist}.

With the charge and current densities at hand  we proceed to the derivation of continuity equations by
taking the equal time limit of the relation~(\ref{eq:Usadel_K}),
\begin{equation}
\lim_{t_1 \to t_2 }\left( \nabla_i J^i({t_1,t_2}) -  \pi \nu [\partial_t, Q]_{t_1, t_2}\right) = 0.
\end{equation}
The diffusive spectral current density ${\bf J}_D(t_1,t_2)$ is non-singular at $t_1 \to t_2$, and hence
\begin{equation}
\lim_{t_1 \to t_2 } \frac 12 {\rm tr}\, \nabla_i J_D^i({t_1,t_2}) \tau_1^K = \partial_{\bf r} {\bf j}_D.  
\end{equation}
However, the CS spectral current is singular,
\begin{equation}
J^z_{\rm CS}(t_1, t_2) = \mp \frac{B_z}{8\pi } \left(
\begin{array}{cc}
\delta(t_1-t_2) & 2 h(t_+, t_-)  \\
0 & - \delta(t_1-t_2)
\end{array}\right),
\end{equation}
and the covariant derivative here has to be evaluated in the similar fashion as above, cf. Eq.~(\ref{eq:[aQ]}). In this way 
we find
\begin{equation}
\lim_{t_1 \to t_2 } \frac 12 {\rm tr}\, \nabla_i J_{CS}^i({t_1,t_2}) \tau_1 = 
\partial_{\bf r} {\bf j}_{\rm SC} \mp \frac{1}{4\pi^2 } B_z E_z,  
\end{equation}
and the continuity equation reflecting the chiral anomaly for each node is given by Eq.~\eqref{eq:NodalContinuity}.

\section{Action of the FCS}

\label{app:FCS_action}

An alternative root to derive the saddle point equations~\eqref{eq:f_n} with the boundary source terms is to consider
the action of the system for the explicit parametrization of the $Q$-matrix~\eqref{eq:Q_flambda}. 
In the quasi-one-dimensional geometry the effective action reduces to the form~\eqref{eq:SeffQuasi1} discussed 
in Sec.~\ref{sec:1dloc}. By representing the field $Q$ in the standard form $Q(x) = T(x) \tau_3 T^{-1}(x)$ --- here
$T(z) = e^{\frac{\lambda(x)}{2}\tau_1} T_0(x)$ and the matrix $T_0(x)$ is specified by Eq.~\eqref{eq:QDist}
with the spatially dependent distribution function $h(x) = 1 - 2 f(x)$, ---
and considering further the limit $\tau_n \to \infty$, one arrives at the following action (given below 
for the node no. 1),
\begin{eqnarray}
 S[f,\lambda] &=&  \sigma_{xx}{\cal A} \int dx d\epsilon\, \Bigl( - f' \lambda' -  f (1-f) \lambda'^{\,2} \Bigr) 
\nonumber \\
 &-&
 N_\phi \int dx d\epsilon \, \left(1/2 - f\right)\lambda'. 
\end{eqnarray}
The 1st term here is also known from the stochastic path integral formulation of the FCS~\cite{Jordan:2004}, while the  
2nd one is a signature of the Weyl semimetal. The Lagrange equations of this action are precisely the
nonlinear equations~(\ref{eq:f_n} -- \ref{eq:lambda_n}).

\bibliography{Bibliography} 

\end{document}